\documentclass[english,twoside,a4paper,10pt]{article}

\usepackage[latin1]{inputenc}
\usepackage[T1]{fontenc}
\usepackage{amsmath}
\usepackage{amsfonts}
\usepackage{graphicx}
\usepackage{subfigure}
\usepackage{a4wide}
\usepackage{amssymb}
\usepackage{fancyhdr}
\usepackage{mathrsfs}
\usepackage[toc,page]{appendix}

\begin{document}

\title{Kinematic reconstructions of extended theories of gravity at small and intermediate redshifts}

\author{ 
Marco Calz\'a$^1$\footnote{E-mail address: marco.calza89@gmail.com
},\,\,\,
Alessandro Casalino$^{1,2}$\footnote{E-mail address: alessandro.casalino@unitn.it}
,\,\,\,
Orlando Luongo$^{3}$\footnote{E-mail address: luongo@lnf.infn.it}
,\,\,\,
Lorenzo Sebastiani$^{1,2}$\footnote{E-mail address: lorenzo.sebastiani@unitn.it}\\
\\
\begin{small}
$^1$
Dipartimento di Fisica, Universit\`a di Trento,Via Sommarive 14, 38123 Povo (TN), Italy
\end{small}\\
\begin{small}
$^2$ TIFPA - INFN,  Via Sommarive 14, 38123 Povo (TN), Italy
\end{small}\\
\begin{small}
$^3$ Laboratori Nazionali di Frascati, Via E. Fermi 40, 00044 Frascati, Italy
\end{small}
}

\date{}

\maketitle

\abstract{In the last few decades, extensions of General Relativity have reached always more attention especially in view of possible breakdowns of the standard $\Lambda$CDM paradigm at intermediate and high redshift regimes. If General Relativity would not be the ultimate theory of gravity, modifying Einstein's gravity in the homogeneous and isotropic universe may likely represent a viable path toward the description of current universe acceleration. We here focus our attention on two classes of extended theories, i.e. the $f(R)$ and $f(R,G)$-gravity. We parameterize the so-obtained Hubble function by means of effective barotropic fluids, by calibrating the shapes of our curves through some of the most suitable dark energy parameterizations, XCDM, CPL, WP. Afterwards, by virtue of the correspondence between the Ricci scalar and the Gauss-Bonnet topological invariant with the redshift $z$, we rewrite $f(R,G)$ in terms of corresponding $f(z)$ auxiliary functions. This scheme enables one to get numerical shapes for $f(R,G)$ and $f(R)$ models, through a coarse-grained inverse scattering procedure. Although our procedure agrees with the simplest extensions of general relativity, it leaves open the possibility that the most suitable forms of $f(R)$ and $f(R,G)$ are rational Pad\'e polynomials of first orders. These approximations seem to be compatible with numerical reconstructions within intermediate redshift domains and match fairly well small redshift tests. }

\section{Introduction}

Unveiling the dark energy evolution from prime principles represents a challenge for modern cosmology \cite{primo}. In particular, at a transition time \cite{secondo}, dark energy dominates over matter though a negative pressure at infrared regimes. This pressure pushes the universe to speed up \cite{terzo}, counterbalancing the action of gravity at large scales. Further, at ultraviolet energies a fully comprehensive approach to quantum gravity is still lacking, leaving the interplay between quantum mechanics and gravity within the domain of pure speculations \cite{kiefer}. Such issues support the timeliness that Einstein's gravity  breaks down at particular scales. This scenario has progressively reached great consensus during the last few decades \cite{Od1,Od2,Od3,Capo1,Capo2,Nojiri:2017ncd}. Several extensions of general relativity have been consequently proposed to account for the cosmic speed up through additional degrees of freedom derived from modified Lagrangians 
\cite{new3,new1,new2,Lazkoz:2018aqk,Casalino:2018tcd,Casalino:2018wnc}.

An intriguing example has been offered by $f(R,G)$ theories, in which the generic function entering the Lagrangian depends upon the Ricci scalar $R$ and the Gauss-Bonnet topological invariant $G$. This treatment explains both early and late-times under the same geometrical scheme, avoiding barotropic additional fluids, dark energy counterparts or ill-defined
scalar fields \cite{Elizalde:2010jx,Makarenko:2012gm,DeLaurentis:2015fea,GBSeba,SantosdaCosta:2018ovq,Odintsov:2018nch}. In particular, the Gauss-Bonnet topological invariant $G$ arises for quantum field theory regularization and is often used within renormalizing fields in curved spacetime, \cite{Birrell}. The Gauss-Bonnet term is motivated since it contributes to trace anomaly as higher-order curvature terms do not vanish. Accounting for a theory in which either $R$ and $G$ contribute into the dynamics exhaust the budget of curvature degrees of freedom required to extend general relativity .

Another widely-used extension of General Relativity is $f(R)$ gravity. Here one considers a single geometric field, i.e. the Ricci scalar, and adopts an analytic function of it as the extended Lagrangians. One of the main disadvantages of both the models lies on postulating the forms of $f(R,G)$ and $f(R)$ which are unknown a priori. Unfortunately, postulating $f(R,G)$ and $f(R)$ would consequently influence the corresponding large-scale dynamics. This does not permit one to reproduce the evolution of the two scenarios and does not enable cosmologists to \emph{disentangle} extensions of general relativity from the standard cosmological model, leading to a severe degeneracy problem.

In this paper, we wonder how to reconstruct the forms of $f(R,G)$ and $f(R)$ functions in a model-independent way. To do so, we presume to rewrite all quantities of interest by means of a single variable, i.e. the redshift $z$. This is possible as one assumes the cosmological principle to hold. Under these hypotheses, we consider a specific Hubble function form in terms of $z$ and build up initial conditions which agree with kinematic requirements, written in a model-independent way. Then we find a $f(z)$ auxiliary function and we frame out the universe evolution in terms of $z$. To do so, we rewrite $f(z)$ as a function of $R$ and $G$, inverting the $R(z)$ and $G(z)$ functions computed with the Hubble function form considered. Finally, we will extend the results at intermediate redshift using an extrapolation approach with the Pad\'e series.

The paper is structured as follows. In Sec. \ref{formalism} we present the theoretical framework, with an emphasis on the dynamics of generic $f(R)$ and $f(R,G)$ theories. In Sec. \ref{sec:reconstruction} we introduce the reconstruction procedure: firstly we list the Hubble function parametrization considered throughout the paper, then we present the numerical approach, among with the results obtained. Finally, in Sec. \ref{sec:extrap} we extend the results of the previous section at intermediate redshifts, employing the Pad\'e series.

\section{Theoretical background\label{formalism}}

We here summarize the principal theoretical requirements necessary to extend general relativity. We focus on $f(R,G)$ and $f(R)$ gravity and we presume the validity of the cosmological principle, assuming the universe to be homogeneous and isotropic. We thus take a  flat  Friedmann-Robertson-Walker (FRW) line element\footnote{The hypothesis of flatness is today debated \cite{ratra1000}. In this work, however, we assume $k=0$ for simplicity, without entering the issue of a non-flat universe. Our results will not be significantly influenced by this choice.}

\begin{equation}\label{FRW}
ds^2=-dt^2+a(t)^2(d{\bf x}^2+ d{\bf y}^2 +d{\bf z}^2)\,.
\end{equation}

\noindent As basic demands suggest, the whole information is encoded in the monotonically-increasing scale factor $a\equiv a(t)$, as inflationary phases ends up. We limit our treatment to fulfill the above requirement at late (dark energy dominated period) and intermediate (standard and dark matter dominated period) universe. We thus presume the concordance $\Lambda$CDM model to well approximate the dynamics at small redshifts. We check possible departures from GR fixing late time boundary conditions over $f(R,G)$ and $f(R)$ through initial settings imposed by kinematics of our models. The scenarios here employed are summarized below.

\subsection{$F(R,G)$ gravity}

The $F(R,G)$-gravity is characterized by the action:
\begin{eqnarray}\label{lagraFRG}
S = \int_\mathcal{M} d^4 x \sqrt{-g} \left[ \frac{F(R, G)}{2}
+{\mathcal{L}}^{\mathrm{(matter)}} \right]\,,
\end{eqnarray}
having physical units of $k_{\mathrm{B}} = c = \hbar = 1$ and Planck's mass set to $8\pi /M_{Pl}^2 =1$. In the picture of Eq. \eqref{lagraFRG}, $g$ is the determinant of the metric tensor $g_{\mu\nu}=g_{\mu\nu}(x^\sigma)$ whereas  $\mathcal{L}^{(matter)}$ is the standard matter Lagrangian and  $\mathcal{M}$ is the 4-dimensional space-time differential manifold.

The function $F(R,G)$ depends upon the Ricci $R$ scalar associated to $g_{\mu \nu}$, and on the Gauss-Bonnet topological invariant $G$, defined by:
\begin{eqnarray}
G \equiv R^2 - 4 R_{\mu \nu} R^{\mu \nu} + R_{\mu \nu \rho \sigma} R^{\mu \nu \rho \sigma}\,,
\end{eqnarray}
where $R_{\mu \nu \rho \sigma}$ and $R_{\mu \nu}$ are the Riemann and Ricci tensors respectively.

If no symmetries or physical constraints are involved, the form of $F(R,G)$ turns out to be \emph{a priori} unknown. Thus, considering the metric in equation \eqref{FRW}, the corresponding modified Friedmann equations, hereafter  \emph{equations of motion} (EOMs), on the FRW background become in general
\begin{eqnarray}
&&  3 H^{2}F_R =\rho +
\frac{1}{2}
\left[ \left( {F}_{R}R+F_G G-F\right)
-6H{\dot{F}}_{R}-24 H^3 \dot F_G
\right]\,,\label{EOM12}
\\
&&-\left( 2\dot H+3H^{2} \right)F_R =p +\frac{1}{2} \Bigl[
-\left( {F}_{R}R-F \right)
+4H{\dot{F}}_{R}+2{\ddot{F}}_{R}\nonumber\\
&&+16 H(H^2+\dot H)\dot F_G
+8H^2 \ddot F_G
\Bigr]\,, \label{EOM22}
\end{eqnarray}
where dots here represent derivatives with respect to the cosmic time $t$ and $H\equiv\dot{a}(t)/a(t)$ is the Hubble function. The EOMs \eqref{EOM12}-\eqref{EOM22} depend on $F\equiv F(R,G)$ and its derivatives, with the additional requirement that $\rho$ and $p$ are the total energy and pressure contents, including baryons\footnote{We here take pressure-less matter and we neglect neutrinos and radiations. For a different perspective over the form of standard matter see Ref.~\cite{luongomuccino}.}, cold dark matter, neutrinos and so forth. From Eqs. \eqref{EOM12}-\eqref{EOM22}, it is easy to see that $ F(R,G)=F(R(t),G(t))$. On the FRW space-time, the invariants $R$ and $G$ take a simple form, which depends on $H$ and its derivative, namely
\begin{subequations}\label{R_and_G_Friedmann_eqs}
\begin{align}
R&=12H^2\left(1+{1\over2}\frac{\dot H}{H^{2}}\right)\,,\\
G&=24H^4\left(1+\frac{\dot H}{H^{2}}\right)\,.
\end{align}
\end{subequations}
We can use the deceleration and jerk parameters, defined as
\begin{subequations}
\begin{align}
q(t) &\equiv-\frac{1}{a H^2}\frac{d^2a}{dt^2}\,,\\
j(t) &\equiv \frac{1}{a H^3}\frac{d^3a}{dt^3}\,,
\end{align}
\end{subequations}
to recast the Hubble function derivatives
\begin{subequations}
\begin{align}
    \dot{H} &= - H^2 \left[ 1 + q(t)\right]\,,\\
    \ddot{H} &= H^3 \left[2 + 3q(t)^2 + j(t)\right]\,,
\end{align}
\end{subequations}
and the equations \eqref{R_and_G_Friedmann_eqs}
\begin{subequations}\label{largo}
\begin{align}
R&=6H^2\left[1-q(t)\right]\,,\\
G&=-24H^4q(t)\,.
\end{align}
\end{subequations}
Further, expanding the luminosity distance in terms of observable quantities \cite{cosmografia}, the present values of $q(t_0)\equiv q_0$ and $j(t_0)\equiv j_0$, can be model independently  measured  \cite{cosmografia2}. The complete set of data is given by
\begin{eqnarray}\label{cosmokinetics}
    H_0&=&74.220^{+5.230}_{-5.080}\,\,\,\text{ Km s$^{-1}$ Mpc$^{-1}$}\,,\\
    q_0&=&-0.615^{+0.272}_{-0.224}\,,\\
    j_0&=&1.030^{+0.722}_{-1.001}\,.
\end{eqnarray}
Considering the experimental values, we can find the value of the Hubble function derivatives today
\begin{subequations}\label{thingsnow1}\begin{align}
\dot{H}(t_0) \equiv \dot{H}_0&= -0.38 H_0^2 \,,\\
\ddot{H}(t_0) \equiv \ddot{H}_0&= 1.18 H_0^4\,,
\end{align}\end{subequations}
and the value of the Ricci tensor and the Gauss-Bonnet topological invariant
\begin{subequations}\label{thingsnow2}\begin{align}
R(t_0) \equiv R_0&=9.69 H_0^2 \,,\label{Riccinow}\\
G(t_0) \equiv G_0&=14.76 H_0^4\,.\label{Gaussnow}
\end{align}\end{subequations}

\subsection{Dynamics of $F(R,G)$ gravity}

To be consistent with the standard concordance model, dubbed the $\Lambda$CDM paradigm, one requires
\begin{equation}
F(R,G)=R+f(R,G)\,,
\label{ast}
\end{equation}
showing the limit to GR as the function $f\equiv f(R,G)$ vanishes or is negligibly small. The $\Lambda$CDM Model corresponds to $f=-2\Lambda$, where $\Lambda$ is the Cosmological Constant.
It is well-known that for pure $f(R)$-gravity we must have,
\begin{equation}
    |f_R| \ll 1\,,\quad
    f_{RR}>0\,.
    \label{cond}
\end{equation}
The first condition avoids substantial corrections to the effective Newton constant of the theory. This can be seen from the modified Friedmann Eqs. \eqref{EOM12}-\eqref{EOM22}, since the canonical $3 H^2$ term (in the first equation) is multiplied by $F_R = 1 + f_R$. If we want this correction, which modifies the effective Newton constant, to be small we should consider a slowly varying $f$ function with respect to $R$. The second condition ensures we do not fall into matter instabilities. This happens because the scalaron mass depends upon $f_{RR}$, so that a negative $f_{RR}$ indicates a negative mass for the scalaron \cite{scalaronz}.

In turn, at infrared scales we thus presume $f(R)$ to weakly evolve with respect to cosmic time. In analogy, if we consider the general case of $f(R, G)$-gravity, we find that the first condition above must be still satisfied, while the second one becomes,
\begin{equation}
9 f_{RR}+6 R f_{R G} +R^2 f_{GG}>0\,, \label{cond2}
\end{equation}
easily satisfied in the simplest case $|R f_{RR}|\,, |R^2 f_{R G}|\,, |R^3 f_{GG}|\ll 1$, as derived in~\cite{monica} for de Sitter space-time, and generalized to a
background with local constant curvature~\cite{fRGcond}.

To simplify our numerical computation, we employ non-dimensional functions, making use of the normalization\footnote{Here, we take $H_0$ as present value of $H$ evaluated at our time, i.e. $t_0$. Although a severe tension occurs \cite{tensione}, this leaves unaltered our final outcomes since all quantities of interest are re-written accordingly to our choice.} $f_G\rightarrow H_0^2 f_G$.
Moreover, we recast Eqs. \eqref{EOM12}--\eqref{EOM22} in terms of a single variable, namely the redshift parameter, defined in terms of the scale factor by $z=-1+\frac{1}{a(t)}$, where we fix $a(t_0)=1$. With this definition in mind, and considering that $a(t)$ is monotonically increasing, we can also rewrite the equations in terms of $z$ instead of using the cosmological time $t$. In particular, the derivatives with respect to time in terms of the redshift using $dz/dt=-(z+1)H$. The procedure to rewrite all quantities in term of the red-shift $z$ has been widely used in several works \cite{proceduracomputazionale} and permits to frame out the shapes of $f(R,G)$ in terms of $z$ only.

It is so possible rewrite  \eqref{EOM12} and \eqref{EOM22} in terms of the only variable $z$, taking in to account  \eqref{ast} and consider the linear combination  \eqref{EOM12}--\eqref{EOM22}:
\begin{equation}
(1+z)\frac{H'}{H}=\frac{\frac{\rho+p}{H^2}+(1+z)\left[A(z)+4H^2 B(z)\right]}
{2+2f_R-8(1+z)H^2f'_G}\,.
\label{equazioneRG0}
\end{equation}
This is the differential equation we will use in order to reconstruct the shape of the $f$ in terms of the red-shift parameter, where the prime index denotes the derivative with respect to the red-shift and $A(z)$ and $B(z)$ read
\begin{eqnarray}
A(z)&=&2f'_R+(1+z)\frac{H'}{H}f'_R+(1+z)f''_R\,,\nonumber\\
B(z)&=&2f'_G+(1+z)\frac{H'}{H}f'_G+(1+z)f''_G\,.\label{equazioneRG02}
\end{eqnarray}

We may rearrange the above equations to enable the modifications to the Hilbert-Einstein action encoded in the $f(R,G)$ function as perfect dark fluid source. This barotropic fluid has the following energy density and pressure
\begin{eqnarray}
\rho_{DE}&=&\frac{1}{2}
\left\lbrace  \left[6H^2-6HH'(z+1)\right]{f}_{R}-f
+6H^2(z+1){f'}_{R}\right.
\nonumber\\&&\left.
+24H^4(z+1)f_G'
\right\rbrace\label{rhoDE}\,,\\
p_{DE}&=&\frac{1}{2}
\left\lbrace - \left[6H^2-2HH'(z+1)\right]{f}_{R}+f
-2H^2(z+1){f'}_{R}\right.
\nonumber\\
\phantom{\frac{1}{2}}&&+2HH'(z+1)^2f'_R+2H^2(z+1)^2f''_R\nonumber\\
\phantom{\frac{1}{2}}&&\left.-8H^2\left[H^2-(z+1)HH'\right]f_G'+8(z+1)^2H^4f_G''
\right\rbrace\label{pDE}\,.
\end{eqnarray}
Now the dark energy Equation of State (EoS) parameter is derived as,
\begin{equation}
 \omega_{DE}=\frac{p_{DE}}{\rho_{DE}}\,,
 \end{equation}
 and the modified Friedmann equations \eqref{EOM12}--\eqref{EOM22} assume the compact form
\begin{eqnarray}
3H^2&=&\rho+\rho_{DE}\,,\label{Fl1}\\
-\left[3H^2-2(z+1)H H'\right]&=&p+p_{DE}\,.\label{Fl2}
\end{eqnarray}
Once the choice of $H$ is made, it is possible to rewrite Eq.~(\ref{equazioneRG0}) in terms of a single unknown function only if
$f(R,G)$ is
 a function of a fixed combination of $R$ and $G$, namely
\begin{equation}
f(R,G)\equiv f(X)\,,\quad X\equiv X(R,G)\,.
\end{equation}
In this way it is possible rewriting \eqref{equazioneRG0}--\eqref{equazioneRG02} in terms $f_X$ as
\begin{equation}
(1+z)\frac{H'}{H}=\frac{\frac{\rho+p}{H^2}+(1+z)\left[A(z)+4H^2 B(z)\right]}
{2+2X_R f_X-8(1+z)H^2\left(f'_X X_G+f_X X_G'\right)}\,,
\label{equazioneRG}
\end{equation}
with
\begin{eqnarray}
A(z)&=&2(X_Rf'_X+X'_R f_X)+(1+z)
\times\nonumber\\&&
\left[\frac{H'}{H}(X_Rf'_X+X'_R f_X)+X_R f''_X+X''_Rf_X+2X'_Rf'_X\right]\,,\nonumber\\
B(z)&=&2(X_Gf'_X+X'_G f_X)+(1+z)
\times\nonumber\\&&
\left[\frac{H'}{H}(X_Gf'_X+X'_G f_X) + X_G f''_X+X''_Gf_X+2X'_Gf'_X\right]\,.
\label{equazioneRGbis}
\end{eqnarray}

The $X$ function adopted in this paper is fixed through physical requirements. In particular, we need an invertible $X$ function, i.e. a $X(z)$ function we can invert in order to find $z(X)=z(X(R,G))$. An invertible function proposal, in addition to the obvious $X=R$ for $f(R)$, is
\begin{equation}
X=\frac{G}{R}\,.
\end{equation}
Finally, this $X$ definition ensures that $f_X$ is a dimensionless quantity.

\section{Reconstructing technique at late times}
\label{sec:reconstruction}

If we choose a functional form for $f$ in terms of $X$, the cosmic evolution can be described by the solutions of the Friedmann-like differential equations \eqref{Fl1}-\eqref{Fl2}. Here, we adopt an alternative strategy.
We consider different parametrizations of the Hubble function, each of which fixes a well defined functional form of $H(z)$, and we solve the differential equation \eqref{equazioneRG} in order to find the functional form of $f$ for values of z spanning from $0$ to $1$.
In other words, we reconstruct the functional form of $f(R, G)$ for a given form of the Hubble function whose constant parameters agrees with the latest experimental results.
Moreover, as we will show in Sec. \ref{sec:extrap}, once the functional form of $f(X)$ is known, we can extrapolate its behavior up to intermediate red-shift data with $z>1$. The only requirements hereafter employed are that dark energy dominates inside $z\leq1$, whereas matter dominates at intermediate red-shifts.
We neglect radiation and neutrino contributions to our puzzle. We also assume pressure-less matter ($P=0$), and we consider the modification of gravity $f(R,G)$ as a source for a \textit{dark fluid} that models the acceleration of the universe, i.e. gives a positive acceleration $\ddot{a}>0$.

This section is structured as follows. The Hubble function parameterizations we employ are listed in the following part \ref{sec:hubble}. In Sec. \ref{sec:numerical} we will further clarify and better motivate the numerical procedure for the reconstruction at small redshift $0\leq z<1$. We will refer to the steps presented here in the whole work. Moreover, in Sec. \ref{sec:error} we will define a procedure to check the goodness of the reconstruction results. Finally, the results are shown respectively for $f(R)$ and $f(R,G)$ in sections \ref{sec:fR} and \ref{sec:fR}.

\subsection{Effective parameterizations of the dark fluid}\label{sec:hubble}

In this section we present the parametrization of the Hubble function we use for the reconstruction procedure. In what follows, we adopt the standard energy density notation in terms of the fractional densities defined as $\Omega_i(z) = \frac{\rho_i(z)}{\rho_\text{tot}(z)}$, where $i=m$ refers to as standard  matter  and cold dark matter, while $i=DE$ stands for the dark fluid component. In particular, if we ignore the contribution of radiation, we obtain
\begin{equation}
\Omega_m(z)=\frac{\rho_m(z)}{3H_0^2}\,,\quad
\Omega_{DE}(z)=1-\Omega_{m}(z)\,,
\label{OMEGA}
\end{equation}
where the matter energy density $\rho_m$ is given by the standard form
\begin{eqnarray}
\rho_m(z) = 3 H_0^2 \Omega_m^0 (1+z)^3\,.
\end{eqnarray}
Here, $\Omega_m^0 \equiv \Omega_m (z=0)$ is the value of the fractional density of standard and dark matter today.

\subsubsection{XCDM parametrization}

A first attempt to enable dark energy to vary is offered by the XCDM scenario. Using Eq. \eqref{Fl1}, we can write the Hubble function evolution as the sum of the standard matter density $\rho_m$ defined above and a term for the energy density of the dark fluid defined by the XCMD parametrization as
\begin{equation}
   \rho_{DE}=3H_0^2\Omega_{DE}^0(1+z)^q\,.
    \label{rhoXCDM}
\end{equation}
Therefore we obtain an equation for the Hubble function
\begin{equation}
\frac{H(z)^2}{H_0^2}=\Omega_m^0 (1 + z)^3 + \Omega_{DE}^0 (z + 1)^{q}\,,\quad q=3(1+\omega_{DE})\,,\label{An1}
\end{equation}
where $\Omega_{DE}^0 \equiv \Omega_{DE} (z=0)$, $q$ is a real coefficient and $\omega_{DE}<-1/3$ is the constant equation of state parameter of the dark energy. The conditions on these parameters are such that the dark fluid models an accelerated expansion. As $q\rightarrow 0$ one recovers the $\Lambda$CDM model.

\subsubsection{Chevallier-Polarski-Linder parametrization}

A further extension of XCDM is given by expanding at first order a varying equation of state parameter through the well-consolidate  Chevallier-Polarski-Linder (CPL) parametrization~\cite{Chevallier:2000qy,Linder:2002et}, which is given by
\begin{equation}
\omega_{DE}=\omega_0+\left(\frac{z}{1+z}\right)\omega_1\,, \label{CPL_omega}
\end{equation}
with $\omega_0\,,\omega_1$ free parameters. Therefore, the dark fluid energy density is given by
\begin{equation}
  \rho_{DE}=3H_0^2\Omega_{DE}^0(z+1)^{3(1+\omega_0+\omega_1)}\text{e}^{-3\omega_1\frac{z}{z+1}}\,, \label{rhoCPL}
\end{equation}
and the Hubble function is
\begin{equation}
    \frac{H(z)^2}{H_0^2}=\Omega_m^0 (z+1)^3+\Omega_{DE}^0 (z+1)^{3(1+\omega_0+\omega_1)}\text{e}^{-3\omega_1\frac{z}{z+1}}\,.\label{An2}
\end{equation}
At our time, since $a(t_0)=1$, $\omega_{DE}<-1/3$. This condition is in agreement with the $\omega_{DE}$ condition in the
XCDM scenario.
Moreover, when $\omega_1=0$, we recover exactly (\ref{An1}) after the identification $\omega_0=\omega_{DE}$.

\subsubsection{Wetterich-redshift parametrization}

The last parameterization we consider is the Wetterich-redshift parametrization (WP), which is defined by the following equation of state:
\begin{equation}
    \omega_{DE} = \frac{\omega_0}{\left[1+\omega_1 \ln (1+z)\right]^2}.
\end{equation}
Thus, the energy density of the dark fluid is
\begin{equation}
\rho_{DE}=   3H_0^2 \Omega_{DE}^0 (z+1)^{3\left[1+\frac{\omega_0}{1+\omega_1 \ln(1+z)}\right]}\,, \label{rhoWP}
\end{equation}
and the Hubble function becomes
\begin{eqnarray}
\frac{H(z)^2}{H_0^2} = \Omega_m^0 (z+1)^3 + \Omega_{DE}^0 (z+1)^{3\left[1+\frac{\omega_0}{1+\omega_1 \ln(1+z)}\right]}.\label{An3}
\end{eqnarray}
When $\omega_0=-1$ and $\omega_1=0$ we get the $\Lambda$CDM model.

\subsection{Reconstruction procedure}\label{sec:numerical}

The reconstruction numerical procedure consists in the following steps:

\begin{enumerate}
\item \label{step1}
We consider a specific form for the Hubble function. Each form is fixed considering the parameterizations listed in the previous section.
\item\label{step2} We solve numerically the differential equation Eq. \eqref{equazioneRG} with respect to the red-shift $z$ for $0<z<1$ (when standard and dark matter and the dark fluid dominate over the other components), imposing suitable initial conditions discussed below. So we obtain an approximate form for $f_X$ as function of $z$ which we linearly fit.
\item\label{step3} We numerically invert X(z) and find $z=z(X)$.
\item\label{step4} We reconstruct the function $f_X$, and therefore $f$, in terms of $X$. In order to make this last step, we take $f_X(z)$ found at step \eqref{step2}, write it in terms of $X$ using $z=z(X)$ found in step \eqref{step3}. Then we integrate in X $f_X(X)$ in order to have an approximate form of $f(X)$ which we fit assuming a specific functional form.
\end{enumerate}

The numerical evaluation has been performed using the values of the cosmological parameters compatible with Planck's results \cite{Ade:2015xua}
\begin{equation}
\Omega_m^0=0.308\qquad \text{and} \qquad \Omega_{DE}^0=0.692\,,\label{obs_value_fract_dens}
\end{equation}
where $\Omega_m^0 \equiv \Omega_m (z=0)$ and $\Omega_{DE}^0 \equiv \Omega_{DE} (z=0)$.

Note that, different Hubble function parametrizations introduce different parameter dependencies for the $f(X)$ function. In fact, the result of the differential Eq. (\ref{equazioneRG}), and in general all quantities analyzed, depend on the parameter used for the parametrizations of $H$. For instance, $F(z)$ and $z(X)$ depend on the parameter $q$ if we use the XCDM parametrization; while using the WP and CPL parametrization, we obtain a dependence on both the two parameters, $\omega_0$ and $\omega_1$. In our numerical evaluations we will only vary one parameter at a time, and fix the others to a reasonable value inferred from the observational data. For instance, in the WP and CPL parametrization we will fix the value of $\omega_0$ and consider $\omega_1$ as the free parameter. A generalization of this approach, with more free parameters, might be considered in future works.

In the following paragraphs we analyze in more details some of the steps above.\\

\paragraph{Step \eqref{step2}: initial conditions.}
The initial conditions we consider are
\begin{equation}
f_X(0)=0\,\qquad \text{and} \qquad f^{\prime}_{X}(0)=0\,,
\end{equation}
namely we require that at the present time where the modification of gravity is dominant the dark energy density is almost a constant. This choice is consistent with the fact that one expects small departures from $\Lambda$CDM Model at the present time.\\

\paragraph{Step \eqref{step4}: $f(R,G)$ proposals.}
In this step we consider $f_X(z)$ found at step \eqref{step2} and rewrite it in terms of $X$ using the function $z=z(X)$ from step \eqref{step3}. An integration step allow us to obtain $f(X)$. Therefore we can infer the shape function $f$ in terms of $X$ once we choose a proposal fitting function. A first reasonable choice for our proposal function $f$ is a low order polynomial expansion
\begin{equation}
f(X)=-2\Lambda\left(1-\frac{g(X)}{2\Lambda}\right)\,,\label{proposal}
\end{equation}
where $g(X)$ is a polynomial function, which in general depends on some constant coefficient, and $\Lambda$ is the constant obtained from the integration of $f_X$. We fix its value to be the one of the cosmological constant, in order to obtain the $\Lambda$CDM in the limit of small modified gravity corrections $g(X) \rightarrow 0$.

The polynomial $g(X)$ must satisfy the following conditions:
\begin{itemize}
\item as previously mentioned, in the limit of $\Lambda$CDM model, $g(X)$ should be equal to zero;
\item at $z=0$ we should have $|g(X)|\ll 2\Lambda$ in order to recover the correct amount of dark energy;
\item for $f(R)$-gravity we must have $|g_R(X)|\ll 1$ and $g_{RR}(X)>0$ in order to satisfy the viability conditions in (\ref{cond}).\\
For $f(X)$-gravity with $X=G/R$, the matter stability condition (\ref{cond2}) turns out to be,
\begin{equation}
f_{XX}\left(\frac{9 G^2}{R^4}-\frac{6 G}{R}+1\right)+\frac{6 f_X}{R}\left(\frac{3G}{R^2}-1\right)>0\,.  \label{cond2X}
\end{equation}
Thus, since for $0.5\lesssim z$ (matter era) one can verify that $G<0$, it is enough to have $f_{XX}>0$ and $f_X<0$, while condition $|f_R|\ll 1$ is still valid.

\end{itemize}

\subsection{Error estimation}\label{sec:error}

Once the reconstruction of $f(X)$ has been completed, we can check the goodness of our results comparing the quantities computed with the $f(X)$ just found with the background functions we started with. For instance we can evaluate the discrepancies between the cosmological evolution predicted by our modified gravity model (defined by the reconstructed $f$), and the cosmological evolution predicted by the corresponding parametrizations of the Hubble function. For this purpose we define the discrepancy function
\begin{equation}
\Sigma_{J} (z)=2\left[\frac{J_\text{par}(z)-J_\text{MG}(z)}{J_\text{par}(z)+J_\text{MG}(z)}\right]\,,\label{Sigma}
\end{equation}
where $J$ is the function of which we are evaluating the discrepancy between the function computed with the reconstructed $f$, $J_\text{MG}(z)$, and the one computed with the evaluation given by the parametrization considered, $J_\text{par}(z)$. Smaller values of this function means better accordance between the reconstruction and the starting setting given by the parametrization considered.

For example if $J_\text{par} (z) = H^2(z)$ is the square of the Hubble function of one parameterization and $J_\text{MG} (z) = H^2_\text{MG}(z)$ is the square of the Hubble function predicted by the modified gravity model, the discrepancy is estimated with
\begin{equation}
\Sigma_{H^2} (z,\omega)=2\left[\frac{H^2(z,\omega)-H_\text{MG}^2(z,\omega)}{H^2(z,\omega)+H^2_\text{MG}(z,\omega)}\right]\,,
\end{equation}
where we add the dependence on the parameter of the parametrization $\omega$.

\subsection{Results for $f(R)$-gravity}\label{sec:fR}

In this chapter we will treat the case of $f(R)$-gravity. We use the general formalism presented in section \ref{formalism}, with $X=R$, and we consider the different forms of Hubble function listed in section \ref{sec:reconstruction}. The numerical procedure for the reconstruction of $f(X)$ here adopted, is explained in section \ref{sec:numerical}. As described in \ref{sec:numerical} in order to obtain $f(X)$ explicitly it is necessary to use a proposal fitting function. In the case of $f(R)$-gravity the same shape fitting function fitts well the numerical samples for all the three considered Hubble function parametrizations, having

\begin{equation}
 g(R) = c_1\omega R+ c_2 \omega^2  (R^2/\Lambda)\,, \label{proposalFR}
\end{equation}

where $\omega$ is the generic parameter of the Hubble function parametrization being $q$ in the case of XCDM and $\omega_1$ in the cases CPL and WP.

\subsubsection{XCDM parametrization}

We begin considering the XCDM parametrization, i.e. we fix the Hubble function given in \eqref{An1} as requested by step \eqref{step1} of the numerical reconstruction procedure presented in section \ref{sec:numerical}. The Planck data \cite{Ade:2015xua} lead to the following viable range for $\omega_{DE}$,
\begin{align}
\omega_{DE}\in[-1.0051;\, -0.961] \quad
\implies \quad q\in[-0.0153;\, 0.117]\,.\label{An1con}
\end{align}
As needed by step \eqref{step4}, we must choose a proposal function. We consider the first two terms of the polynomial expansion of $g(R)$ as defined in Eq. \eqref{proposal},
\begin{equation}
f(R)=-2\Lambda
\left(1- \frac{c_1q R+ c_2 q^2  (R^2/\Lambda)}{2\Lambda}\right)\,.
\label{Mod1}
\end{equation}
By using our numerical reconstruction we are able to reconstruct the expansion of $f(R)$ as a function of $z$ within the range $[0,1]$, i.e. we perform the step \eqref{step2} of the numerical reconstruction procedure. The next two steps provide the coefficient $c_{1,2}$ of the proposal function, whose values are
\begin{equation}
c_1=-0.09\quad\text{and}\quad c_2= 0.054\,.\label{Mod1_param}
\end{equation}
Since $qc_1\,, q^2 c_2 /\Lambda \ll 1$ and $c_2>0$, both conditions (\ref{cond}) are satisfied. Furthermore,
for $q=0$ we get the $\Lambda$CDM Model with $f(R)=-2\Lambda$. We also note that $q^2 c_2$ is extremely small, and this justifies the truncation of terms with order of $R$ higher than two in our expansion of $g(R)$.

\begin{figure}[!h]
\includegraphics[width=0.5\textwidth]{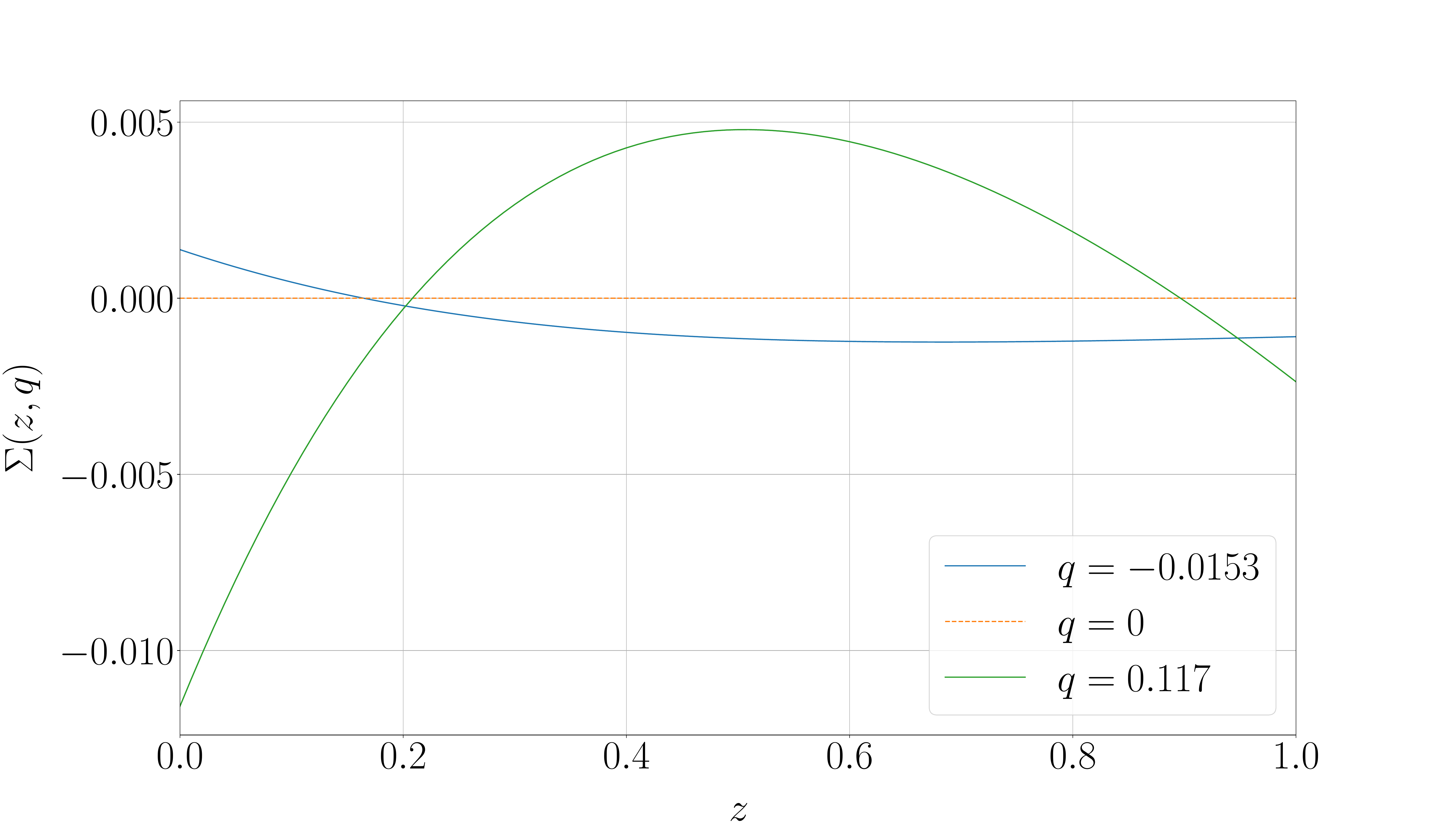}
\includegraphics[width=0.5\textwidth]{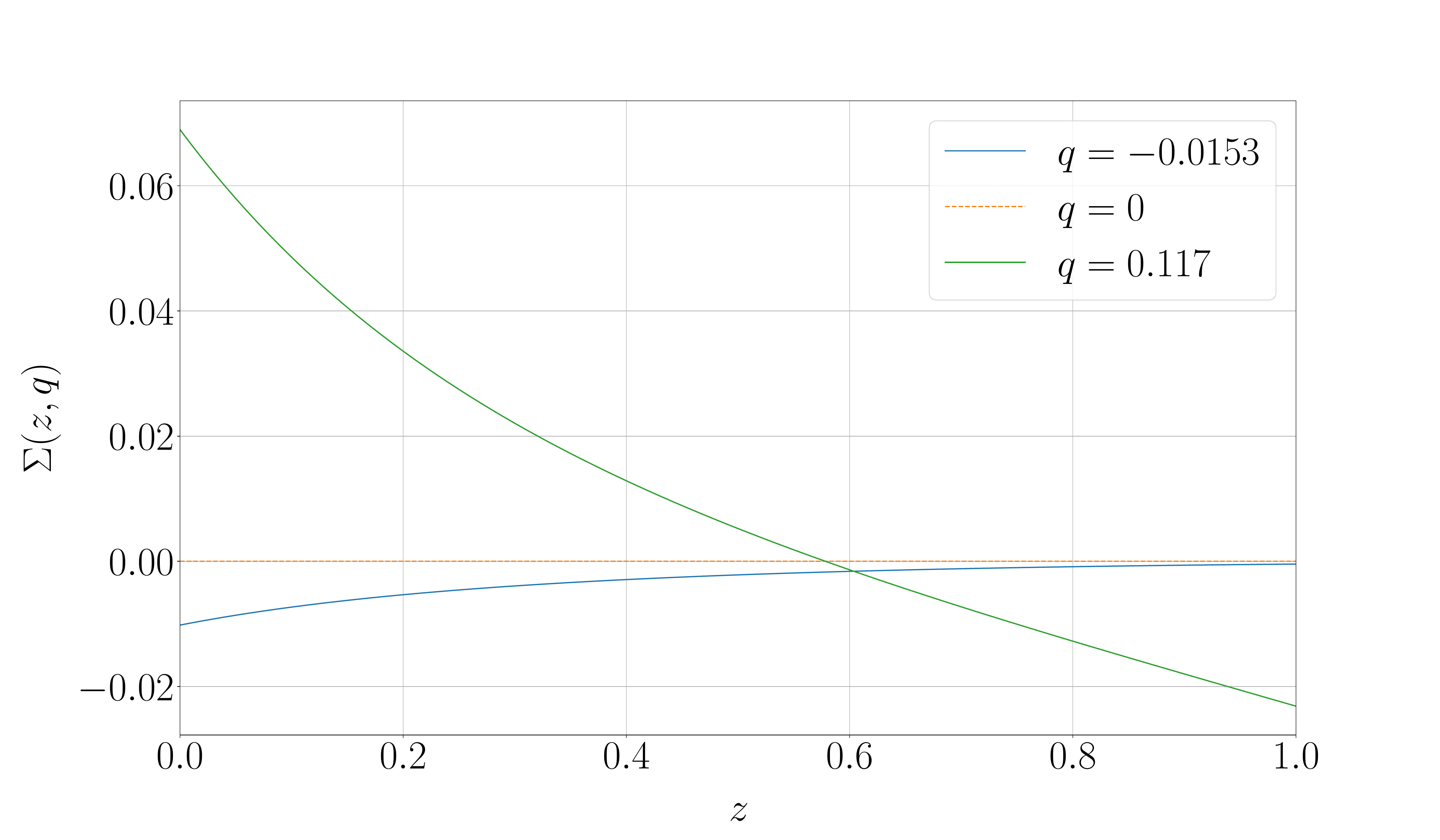}
\caption{Plots of the discrepancy function $\Sigma$, as defined in Eq. \eqref{Sigma}, with $J=H^2$ (a), and with $J=(z+1)H'/H$ (b), for the XCDM parametrization in the case of $f(R)$-gravity.
\label{Fig1}}
\end{figure}

In order to estimate the goodness of the reconstruction, in Fig.~\ref{Fig1}(a) we plot the $\Sigma$ function, as defined in equation \eqref{Sigma}, related to the square of the Hubble function $\Sigma_{H^2}$, for the values of $q$ at the extremes proposed in equation \ref{An1con}. In other words, we evaluate the discrepancy between the value of $H^2$ computed with the reconstructed $f(R)$ function \eqref{Mod1}, with $c_1$ and $c_2$ parameters as in Eq. \eqref{Mod1_param}, and the one computed directly with the Hubble from the parametrization, Eq. \eqref{rhoXCDM}. We note that this error is smaller than $1\%$ within both the redshift and the parameter $q$ ranges considered. Analogously $\Sigma_{(z+1)H'/H}$ is shown in Fig.~\ref{Fig1}(b). In this case, the discrepancy is smaller than $6\%$. For all $q$ in between the extreme values, the discrepancy is always lower.

\begin{figure}[!h]
\includegraphics[width=0.5\textwidth]{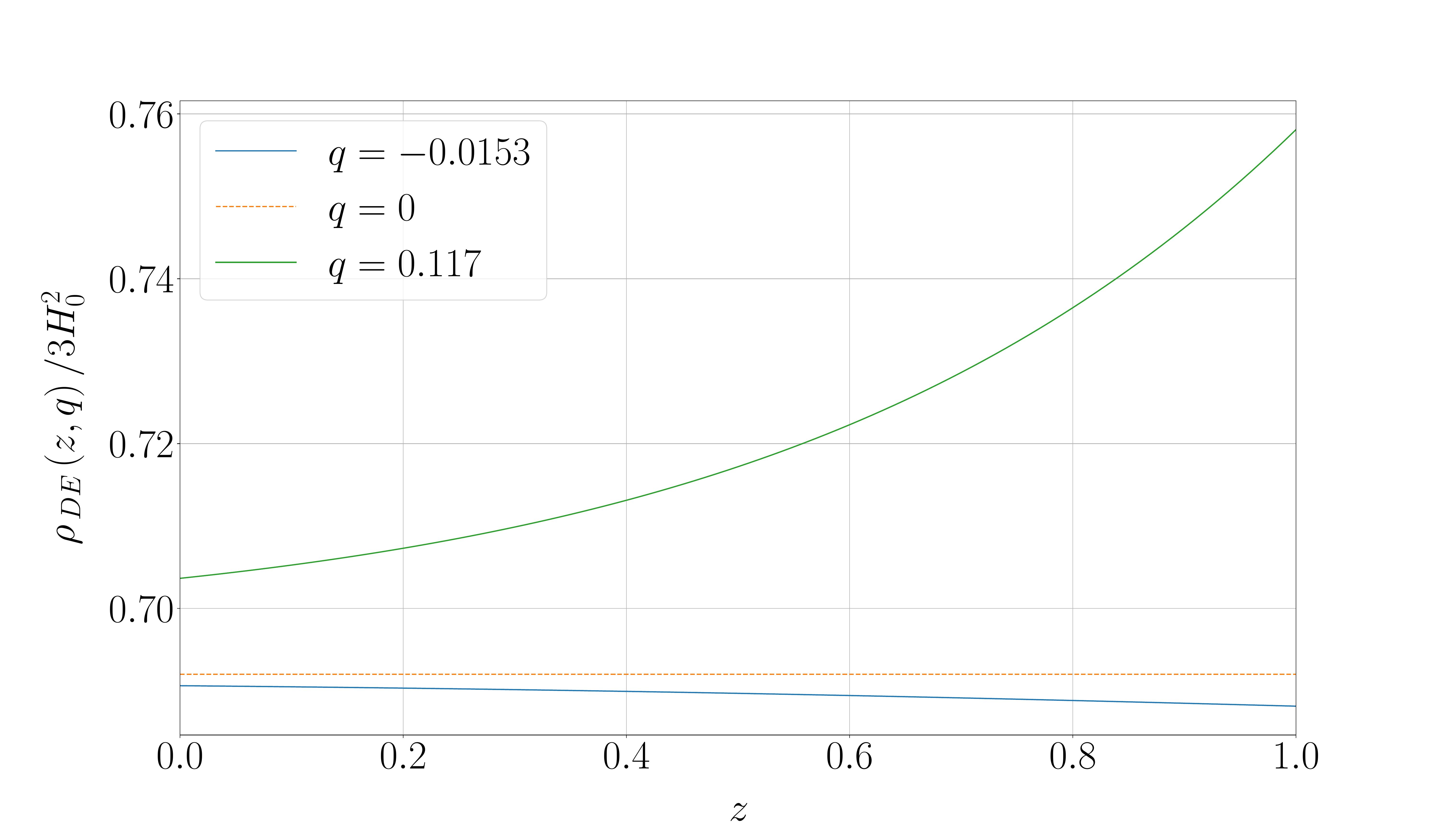}
\includegraphics[width=0.5\textwidth]{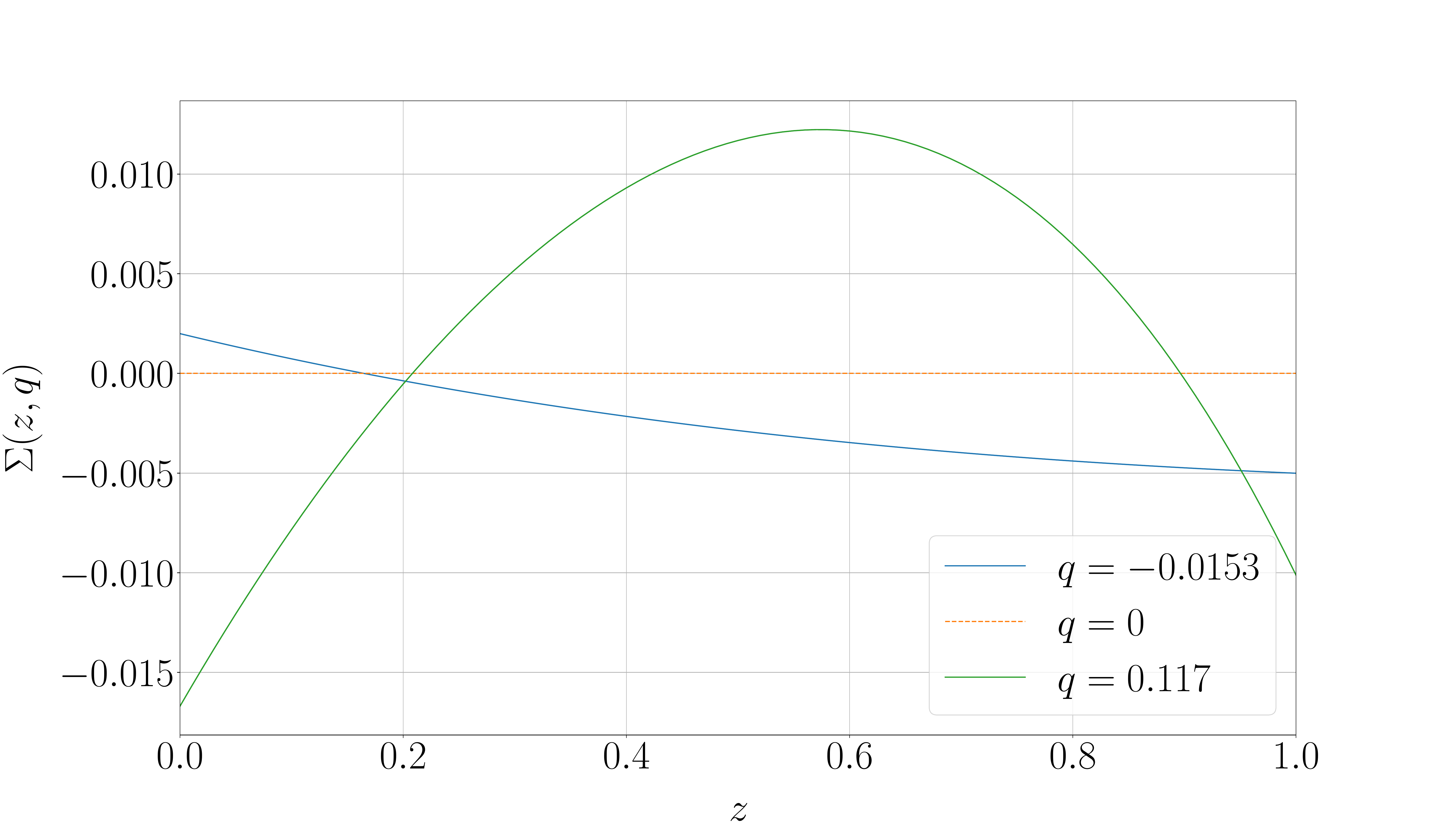}
\caption{The effective energy density from modified gravity, computed with Eq. (\ref{rhoDE}) (a), and its discrepancy $\Sigma$, as defined in Eq. \eqref{Sigma}, against the XCDM parametrization density Eq. \eqref{rhoXCDM} (b), in the case of $f(R)$-gravity.
\label{Fig2}}
\end{figure}

In Fig.~\ref{Fig2}(a) we plot the renormalized energy density of the dark fluid $\rho_{DE}/(3H_0^2)$, as a function of the redshift and the parametrization parameter $q$, computed using Eq. (\ref{rhoDE}).
We note that the evolution mimics the one of a quintessence fluid when $q>0$, i.e. the energy density of the quintessence dark fluid grows up with the red-shift,
and the one of a phantom fluid when $q<0$, which is characterized by an energy density of the phantom dark fluid that goes down with the red-shift. Note that for $z=0$
we obtain $\rho_{DE}/(3H_0^2)\simeq 0.692 \equiv \Omega_{DE}^0$, which is, as expected, the observational value considered in Eq. \eqref{obs_value_fract_dens}.
Finally, the plot in Fig.~\ref{Fig2}(b) shows the discrepancy $\Sigma_{\rho_{DE}}$
between the energy density of the
XCDM parametrization, Eq. \eqref{rhoXCDM},
and the effective energy density of
modified gravity computed with Eq. (\ref{rhoDE}), i.e. the one shown in Fig.~\ref{Fig2}(b). The error is smaller than $1.5\%$, confirming the accordance between the reconstruction and the parametrization functions we started from.

\subsubsection{Chevallier-Polarski-Linder parametrization}

\begin{figure}[!h]
\includegraphics[width=0.5\textwidth]{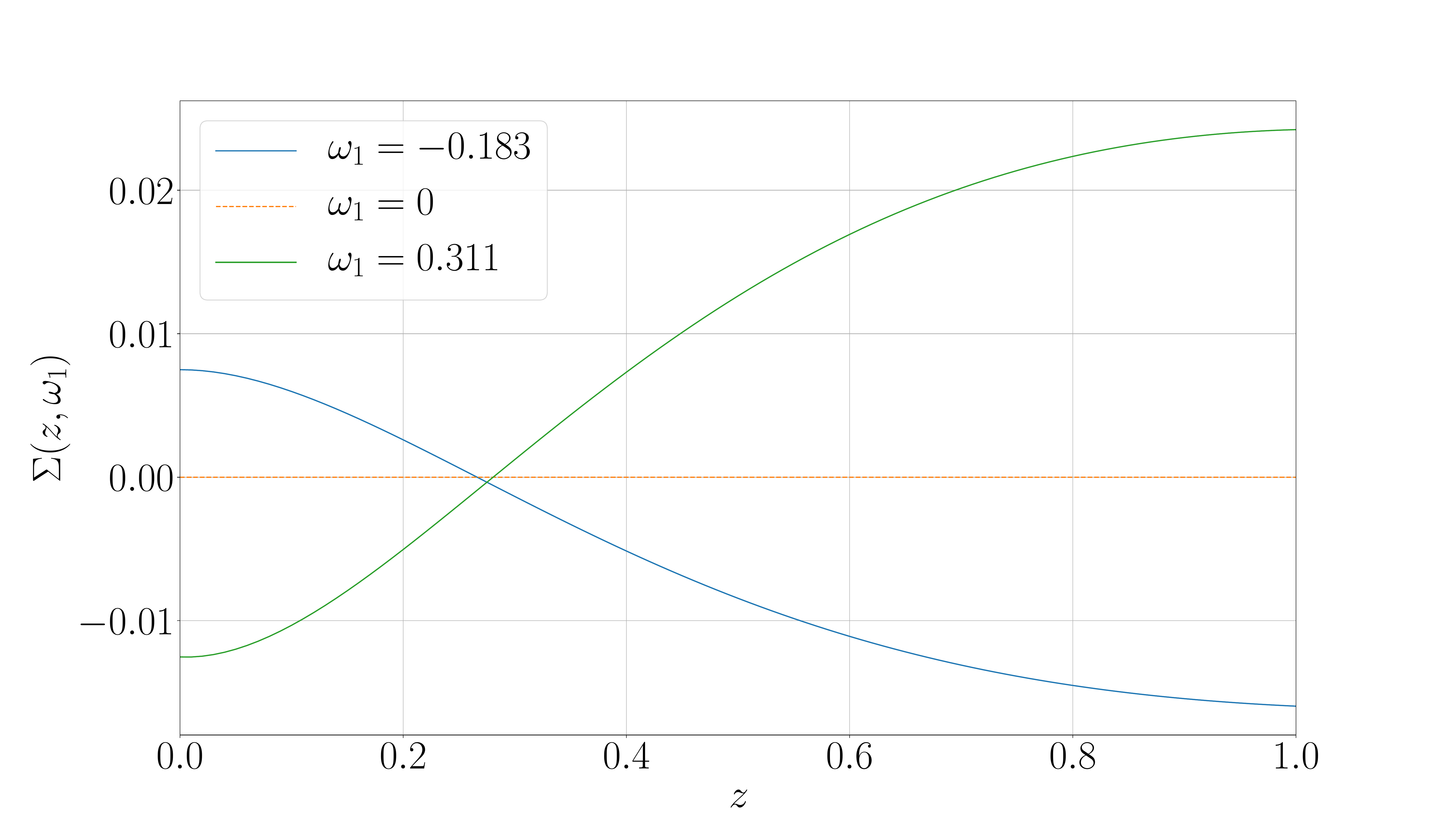}
\includegraphics[width=0.5\textwidth]{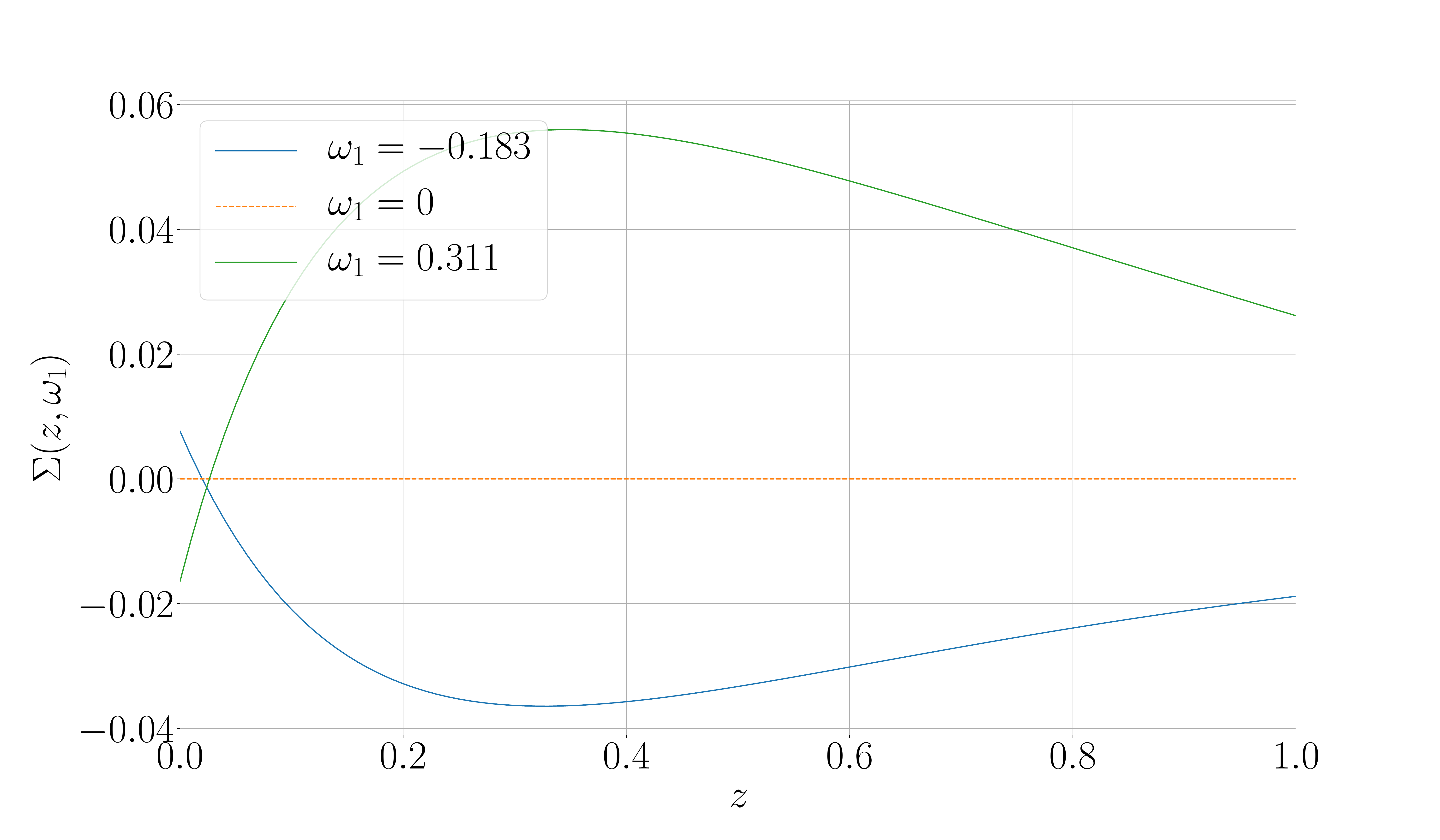}
\caption{Plots of the discrepancy function $\Sigma$, as defined in Eq. \eqref{Sigma}, with $J=H^2$ (a), and with $J=(z+1)H'/H$ (b), for the CPL parametrization in the case of $f(R)$-gravity.
\label{Fig3}}
\end{figure}

In this section we present the result for the CPL parametrization, which has an Hubble function as in Eq. \eqref{An2}. As already mentioned in section \ref{sec:numerical}, we want to vary only one parameter of the parametrization. In order to do so, we fix $\omega_0=-1$, while we consider $\omega_1$ to be in the range as in equation (\ref{omegaCPL}). In other words, we consider the constant part of $\omega$ to be the one of $\Lambda$CDM  and the term $\omega_1$ a correction to this value that depends on the redshift $z$ as shown in the CPL parametrization equation \eqref{CPL_omega}. We consider the range of values of the parameter $\omega_1$ as shown in \cite{Sola:2016hnq}, that is
\begin{equation}
-0.183<\omega_1<0.311\,.\label{omegaCPL}
\end{equation}
For this parametrization we choose a proposal function (as needed by step \eqref{step4} of the reconstruction procedure) equal to the previous one
\begin{equation}
f(R)=-2\Lambda\left(1-{\frac{c_1\, \omega_1 R +c_2 \omega_1^2 (R^2/\Lambda)}{2\Lambda}}\right)\,,
\label{Mod2}
\end{equation}
up to a renaming of the parametrization parameter $\omega_1$. Performing the reconstruction procedure, we obtain as the values of the proposal function parameters
\begin{equation}
c_1=-0.042\quad \text{and} \quad c_2=0.0031.    \label{Mod2_param}
\end{equation}
The discrepancies $\Sigma$, defined in Eq. \eqref{Sigma}, on $H^2$ and on $(z+1)H'/H$
between our reconstructed modified gravity model, defined by the reconstructed $f(R)$ function in Eq. \eqref{Mod2} with parameters \eqref{Mod2_param}, and the starting CPL parametrization setting,
are plotted in Fig.s~\ref{Fig3}(a) and \ref{Fig3}(b), respectively.
The error on the square of the Hubble function is smaller than $2.5\%$, while the one on the ratio $(z+1)H'/H$ reaches at most the $5\%$.

\begin{figure}[!h]
\includegraphics[width=0.5\textwidth]{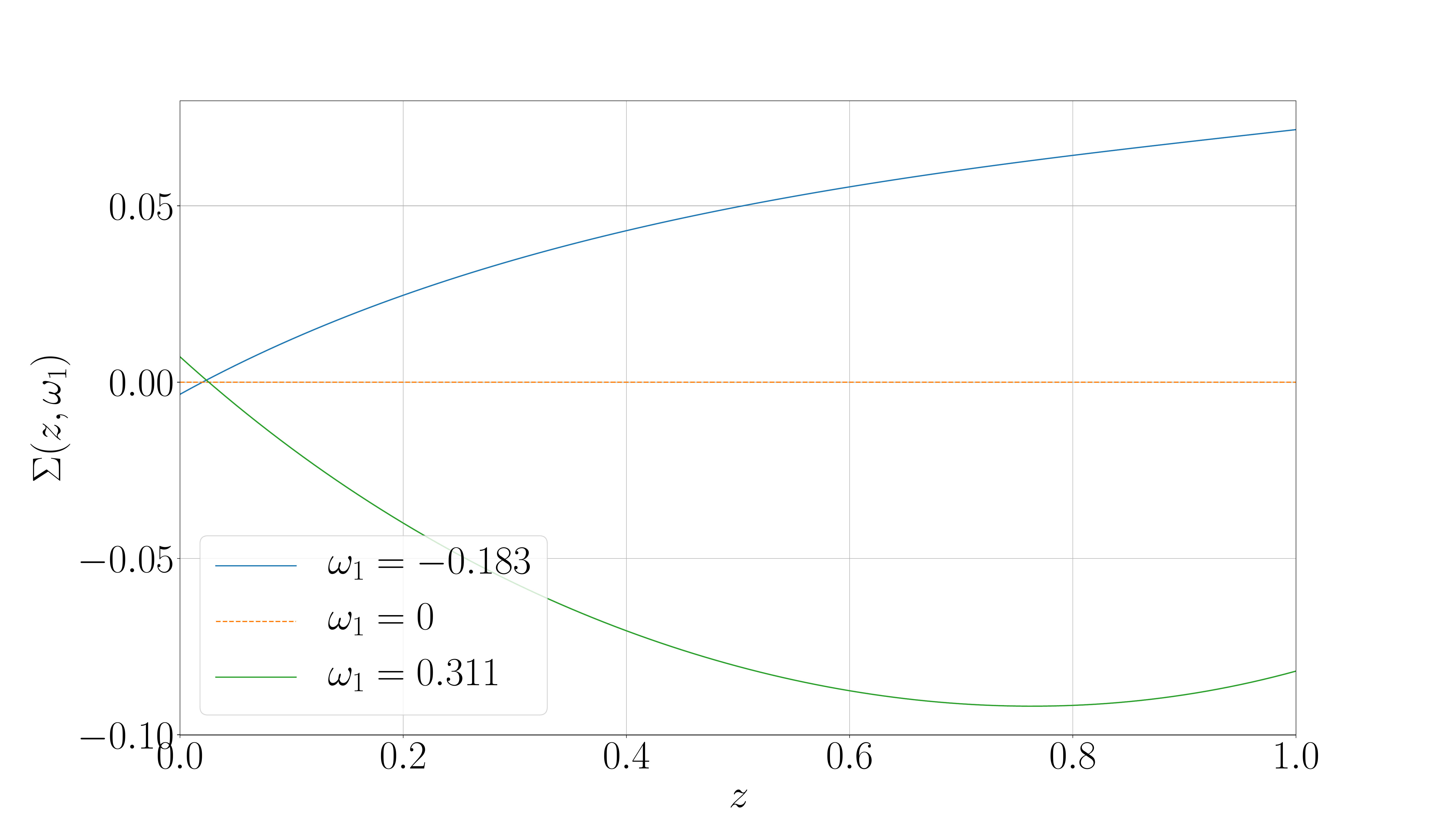}
\includegraphics[width=0.5\textwidth]{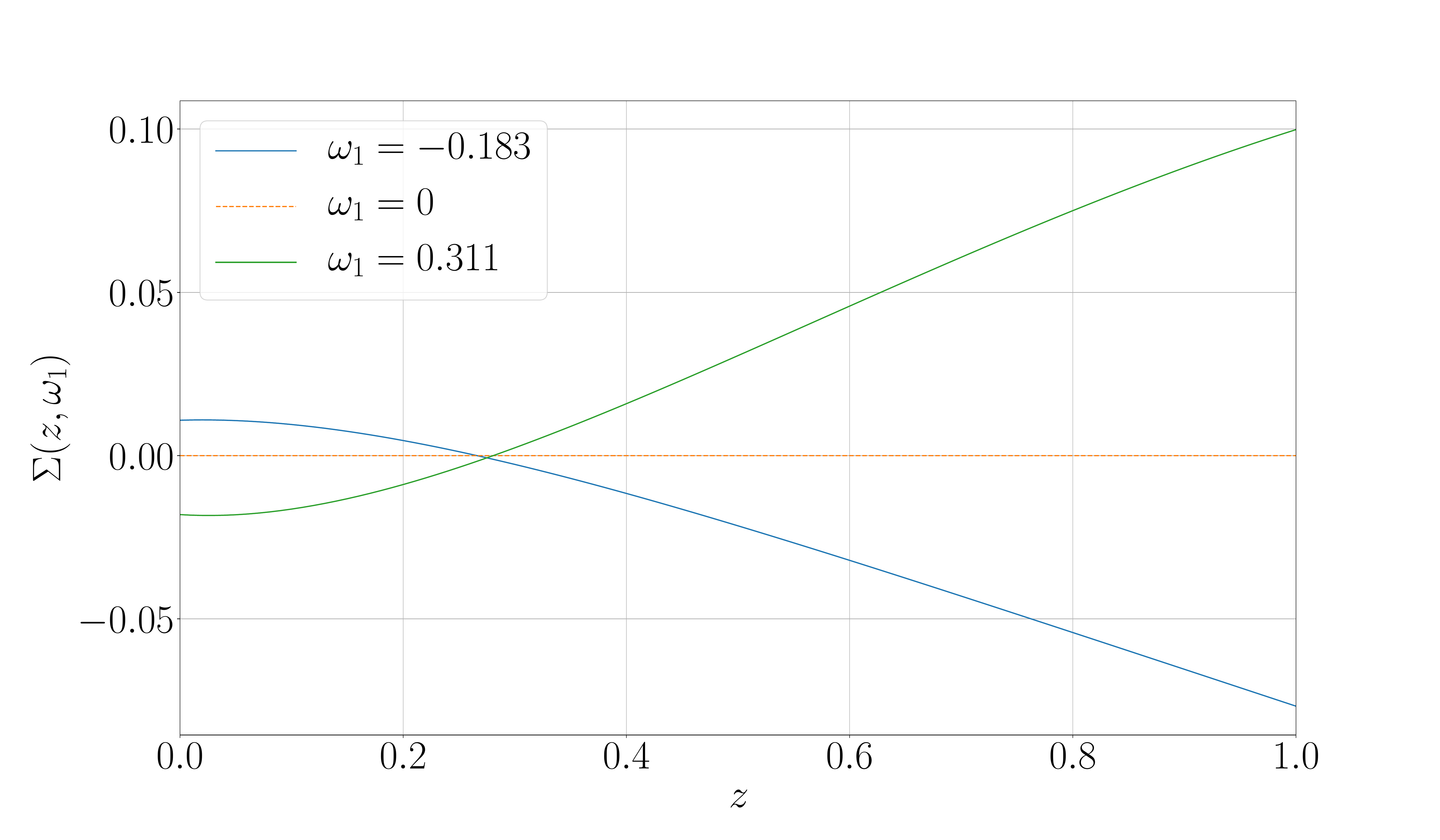}
\caption{The discrepancy $\Sigma$, as defined in Eq. \eqref{Sigma}, with $J=\rho_{DE}/(3H_0^2)$  (a) and with the equation of state parameter $J=\omega_{DE}$ (b), for the CPL parametrization in the case of $f(R)$-gravity.
\label{Fig4}}
\end{figure}

In Fig.~\ref{Fig4}(a) we plot the discrepancy between the energy density of the dark fluid of CPL parametrization \eqref{rhoCPL}
and the effective energy density from the reconstructed modified gravity model, computed with Eq. (\ref{rhoDE}). The plot shows an accordance up to an error of $10\%$, reached at high redshift and for large values of $\omega_1$. Moreover, the discrepancy $\Sigma$ on $\omega_{DE}$, as shown
Fig.~\ref{Fig4}(b), is always smaller than $10\%$. Again, all $q$ in between the extremes, the $\Sigma$ function values is always smaller.

\subsubsection{Wetterich-redshift parametrization}

\begin{figure}[!h]
\includegraphics[width=0.5\textwidth]{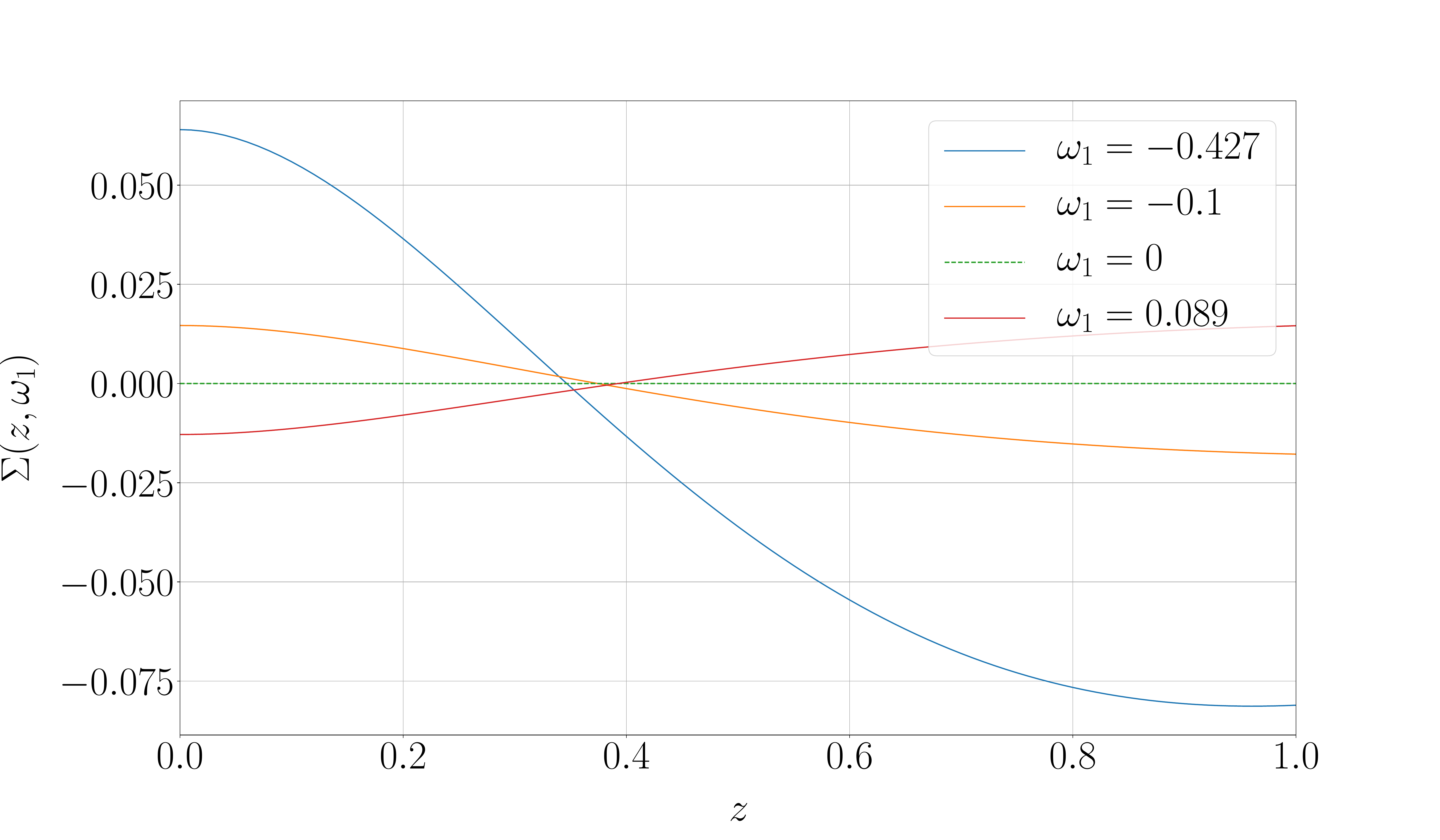}
\includegraphics[width=0.5\textwidth]{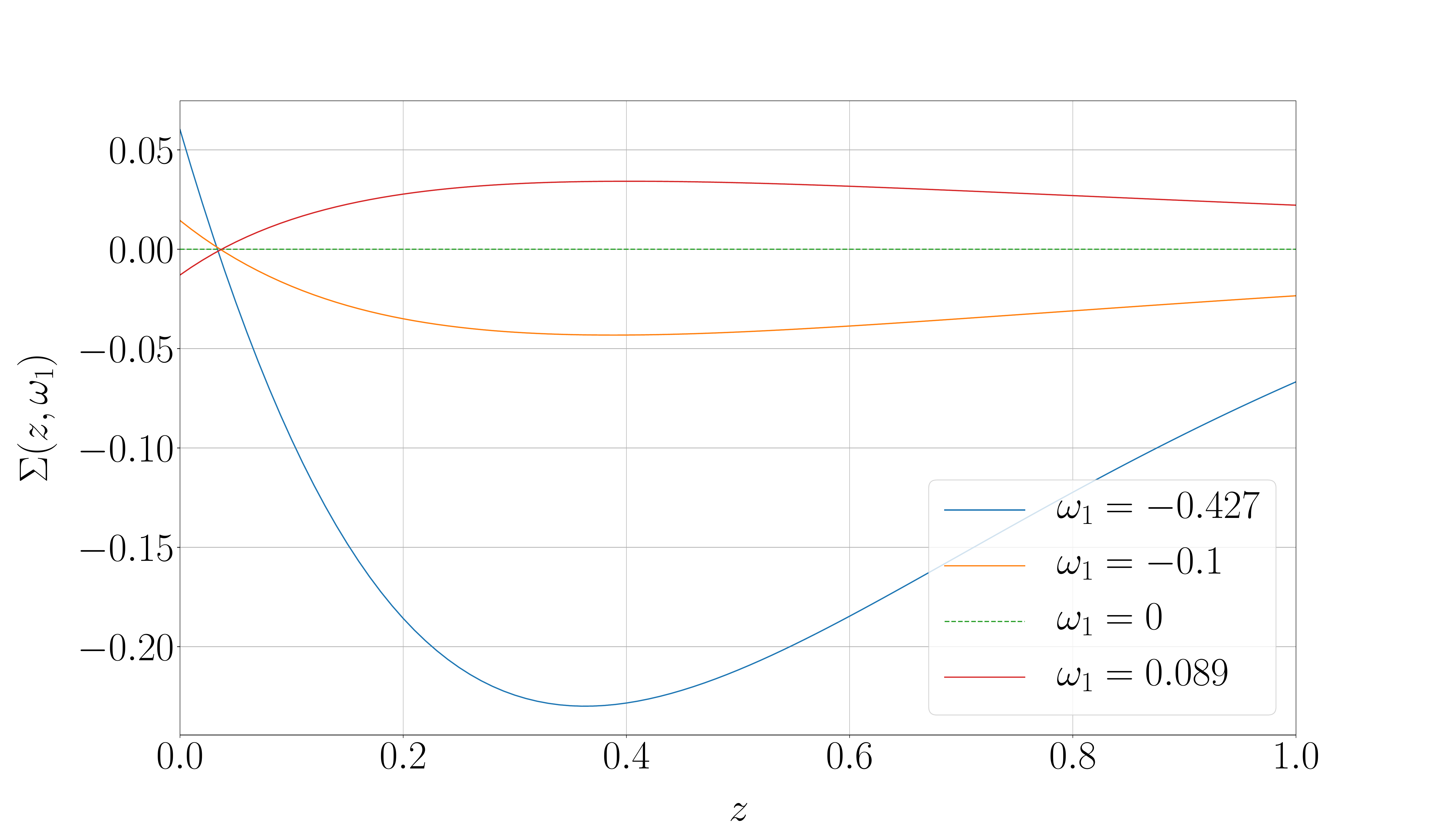}
\caption{Plots of the discrepancy function $\Sigma$, as defined in Eq. \eqref{Sigma}, with $J=H^2$ (a), and with $J=(z+1)H'/H$ (b), for the WP in the case of $f(R)$-gravity. Note that the case with $\omega_1=-0.427$ suffers of high propagation error.
\label{Fig5}}
\end{figure}

The last parametrization we consider is the WP, with an Hubble function defined in Eq. (\ref{An3}). We fix again $\omega_0=-1$, while $\omega_1$ is varied in the range
\begin{equation}
-0.427<\omega_1<0.089\,.\label{om1WP}
\end{equation}
This values interval comes from \cite{WPintervall}.
We define also here the proposal function, as required by step \eqref{step4}. We choose the same function as in the previous parametrizations, which is
\begin{equation}
f(R)=-2\Lambda\left(1-{\frac{c_1\, \omega_1 R +c_2 \omega_1^2 (R^2/\Lambda)}{2\Lambda}}\right)\,.
\label{Mod3}
\end{equation}
Applying the reconstruction procedure presented in section \ref{sec:numerical}, we obtain the values of
\begin{equation}
c_1=-0.15\quad \text{and} \quad c_2=3.9\times 10^{-10}.
\end{equation}
The discrepancies $\Sigma$, defined in Eq. \eqref{Sigma}, of $H^2$ and $(z+1)H'/H$,
between our reconstructed modified gravity model, defined by the $f(R)$ function, and the starting WP setting,
are plotted in Figs.~\ref{Fig5}(a) and \ref{Fig5}(b), respectively. The error on $H^2$ is smaller than $6\%$, while the error on $(z+1)H'/H$ may reach $20\%$ for large and negative values of $\omega_1$.

\begin{figure}[!h]
\includegraphics[width=0.5\textwidth]{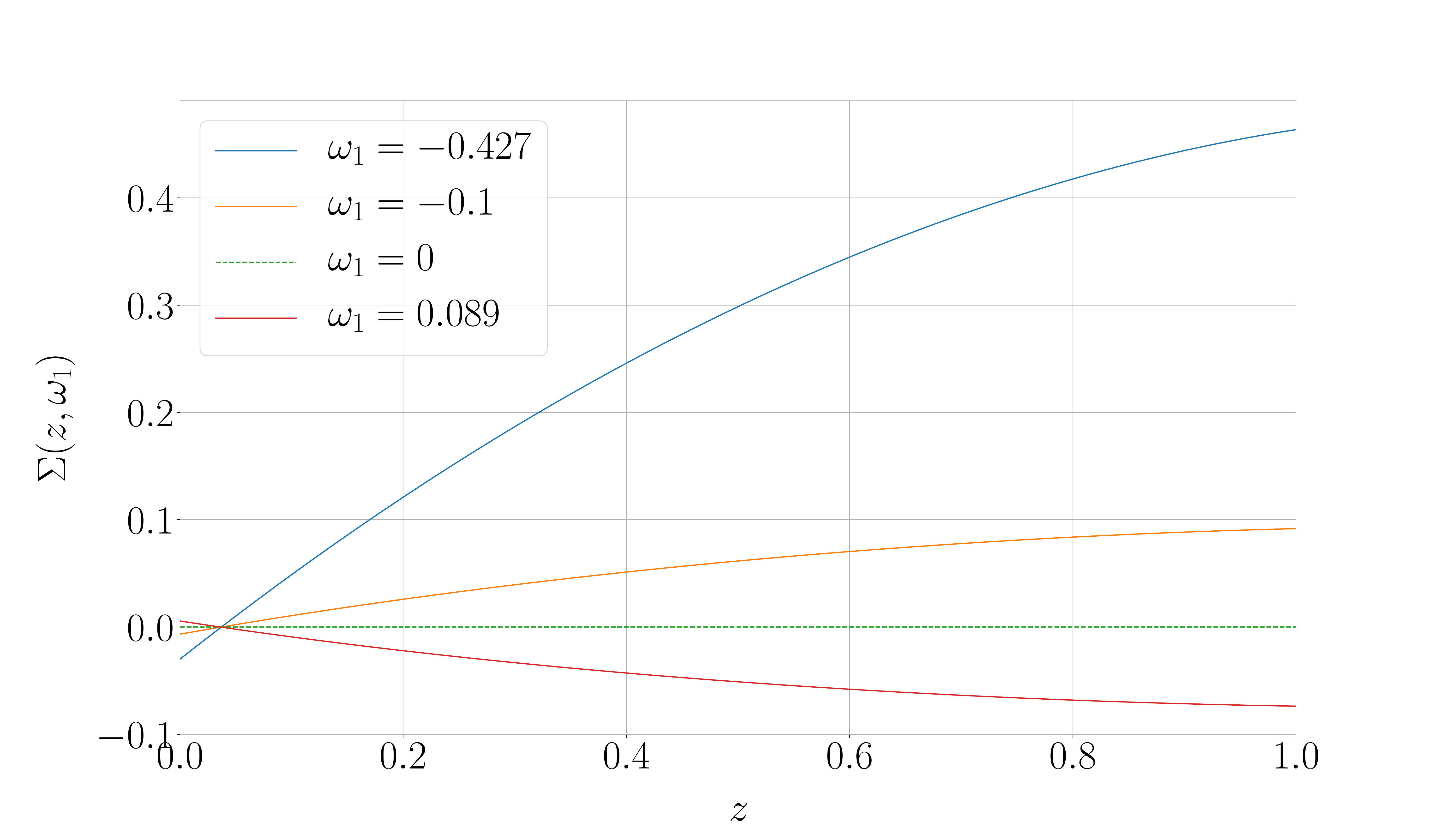}
\includegraphics[width=0.5\textwidth]{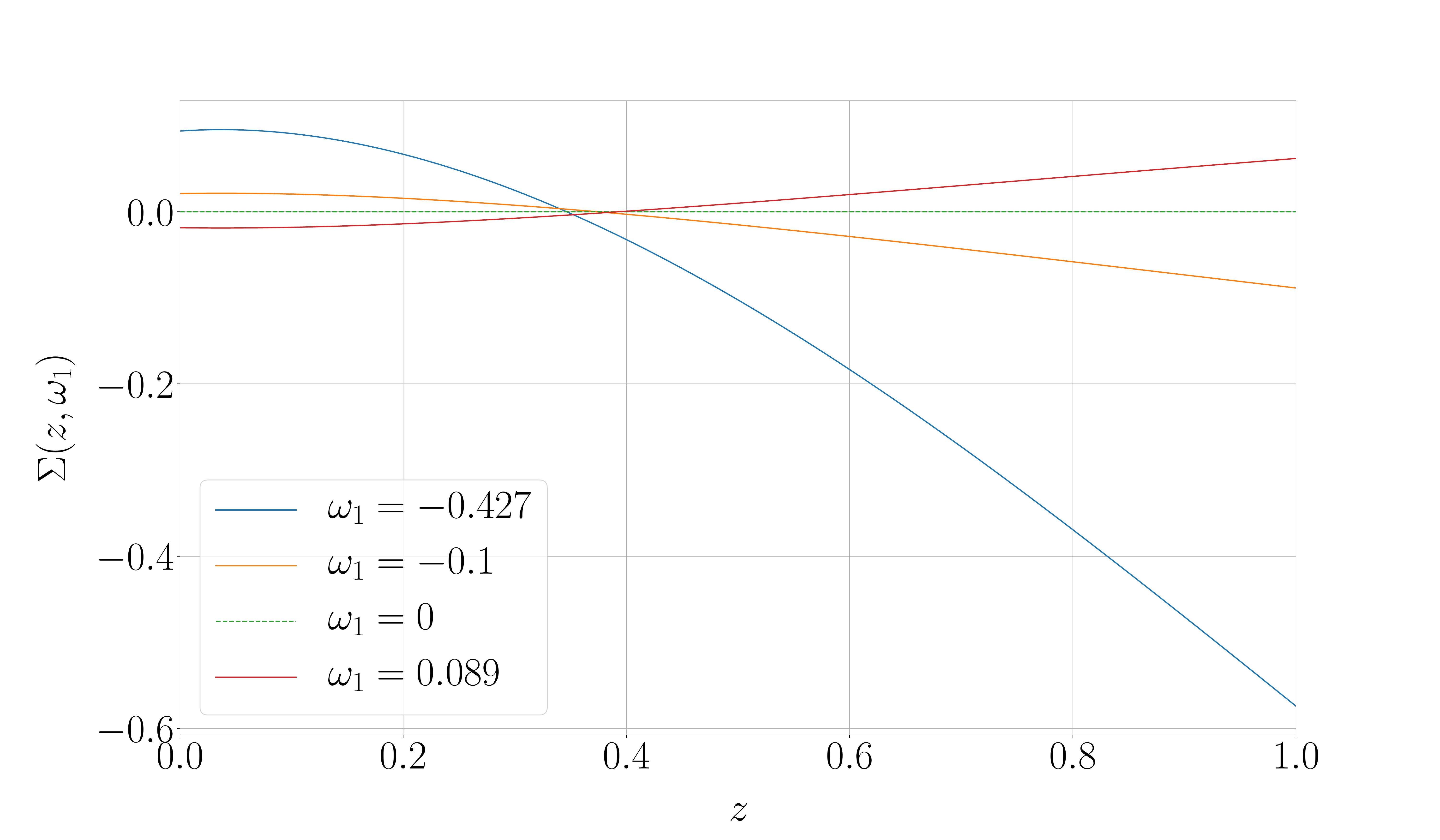}
\caption{The discrepancy $\Sigma$, as defined in Eq. \eqref{Sigma}, with $J=\rho_{DE}/(3H_0^2)$  (a) and with the equation of state parameter $J=\omega_{DE}$ (b), for the WP in the case of $f(R)$-gravity.
\label{Fig6}}
\end{figure}

In Fig.~\ref{Fig6}(a) we plot the discrepancy between the energy density
of the dark fluid of the WP, defined in Eq. \eqref{rhoCPL}, and the effective energy density from the reconstructed modified gravity model. We see that the errors are smaller than $5\%$ in the considered ranges. However, the discrepancy on the effective equation of state parameter $\omega_{DE}$ in Fig.~ \ref{Fig6}(b) shows an error of $> 40 \%$ for large and negative values of $\omega_1$.

From the last consideration we can conclude that, using an $f(R)$ model and a proposal function in the form of Eq. (\ref{Mod3}), we can not reconstruct a viable $f(R)$ using WP evolution with sufficiently low error. However, if we restrict the range of $\omega_1$ we can still use the reconstruction model. In fact, within the range of the parameter
\begin{equation}
   -0.1\lesssim\omega_1<0.089\,,
\end{equation}
the error are restrained under $10\%$. This situation might be improved choosing a more suitable proposal function.

\subsection{Results for $f(R,G)$-gravity}\label{sec:fRG}
In this chapter we will treat the case of $f(R,G)$-gravity in the particular case $X = X(R,G)= \frac{G}{R}$ (see end of section \ref{formalism}, where we explain the viability of this form for $X$ in the contest of reconstruction methods). We consider the different forms of Hubble function listed in section \ref{sec:reconstruction}. The numerical procedure for the reconstruction of $f(X)$ here adopted, is explained in section \ref{sec:numerical}.
As described in \ref{sec:numerical} in order to obtain $f(X)$ explicitly it is necessary to use a proposal fitting function. In the case of $f(R,G)$-gravity the same shape fitting function fitts well the numerical samples for all the three considered Hubble function parametrizations, having

\begin{equation}
 g(X) = c_1\omega X+ c_2 \omega  (X^2/\Lambda)\,, \label{proposalFRG}
\end{equation}

where $\omega$ is the generic parameter of the Hubble function parametrization being $q$ in the case of XCDM and $\omega_1$ in the cases CPL and WP.

\subsubsection{XCDM parametrization}

We begin considering the XCDM parametrization of the Hubble function given in (\ref{An1}). The viable range from the Plank data ~\cite{Ade:2015xua} is given in Eq. (\ref{An1con}). Using again the numerical reconstruction procedure presented in section \ref{sec:numerical}, we are able to reconstruct the expansion of such $f(R,G)$ for the value of $z$ within the range $[0,1]$. In order to do so, we choose, as requested by step \eqref{step4} of the numerical reconstruction procedure, the following proposal function
\begin{equation}\label{Mod1_fRG}
f(R, G)=-2\Lambda\left(1-{\frac{c_1\, q X +c_2 q (X^2/\Lambda)}{2\Lambda}}\right)\,,\quad X=\frac{G}{R}\,.
\end{equation}
The values of the parameters $c_1\,,c_2$ are obtained fitting the numerical sample using  \eqref{Mod1_fRG}
\begin{equation}
c_1=0.11\quad \text{and} \quad c_2=-0.016.
\end{equation}
Note that in step \eqref{step4} we also listed the requirements that $g(X)$ must satisfy. By taking into account Eq.~(\ref{cond2X}) we immediately see that only the negative values of $q$, which corresponds to phantom dark energy values are acceptable. For this reason, the acceptable range of the parameter $q$ turns out to be,
\begin{equation}
-0.0153<q<0\,.
\end{equation}

\begin{figure}[!h]
\includegraphics[width=0.5\textwidth]{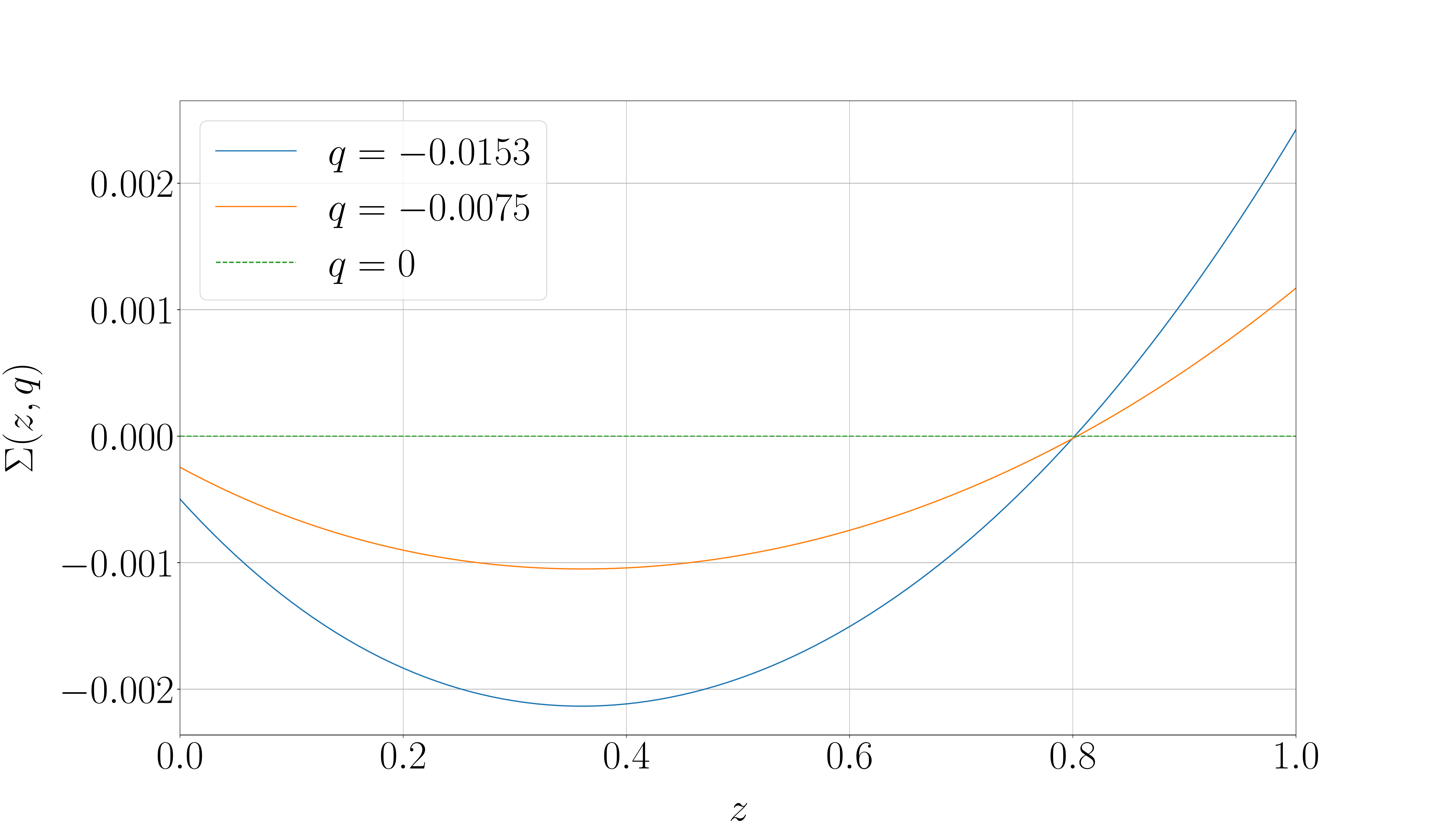}
\includegraphics[width=0.5\textwidth]{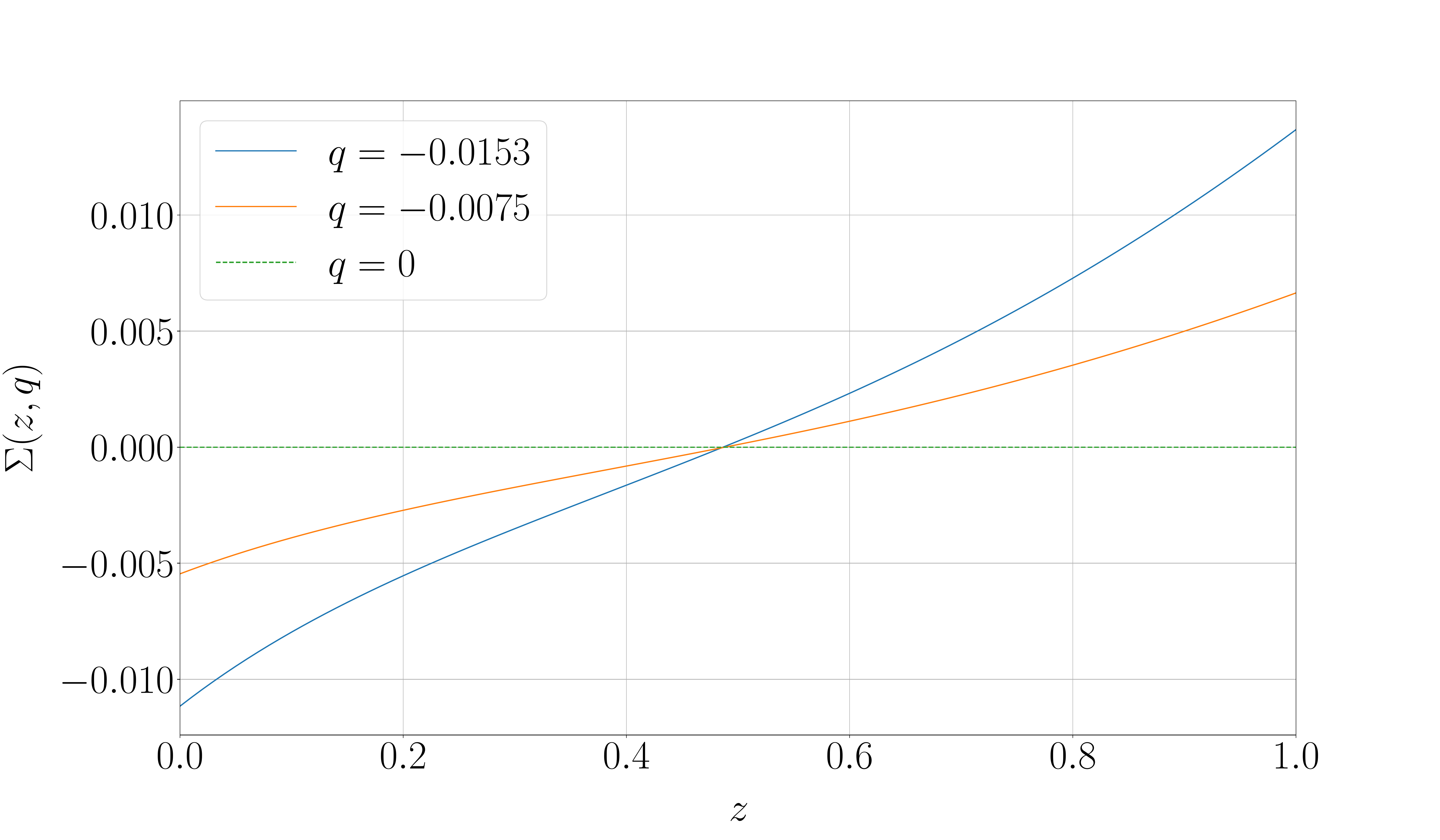}
\caption{Plots of the discrepancy function $\Sigma$, as defined in Eq. \eqref{Sigma}, with $J=H^2$ (a), and with $J=(z+1)H'/H$ (b), for the XCDM parametrization in the case of $f(R,G)$-gravity.
\label{Fig7}}
\end{figure}

In Fig.s \ref{Fig7}(a) and \ref{Fig7}(b) we plot the discrepancies on $H^2$ and on $(z+1)H'/H$ between the XCDM parameterization and our reconstructed modified gravity model. The discrepancy on $H^2$ is smaller than $0.3\%$, while the discrepancy on $(z+1)H'/H$ is smaller than $1\%$.

\begin{figure}[!h]
\includegraphics[width=0.5\textwidth]{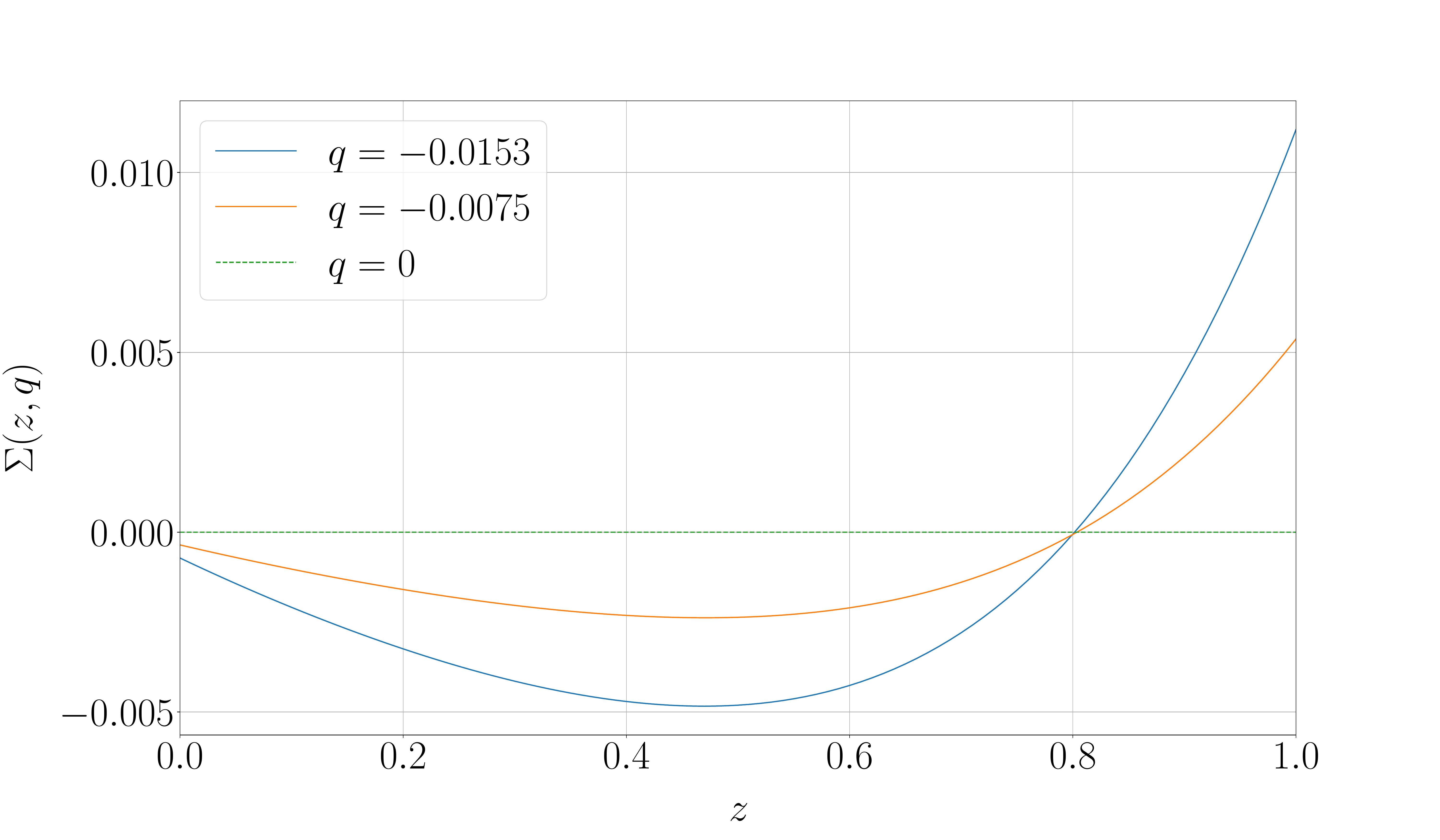}
\includegraphics[width=0.5\textwidth]{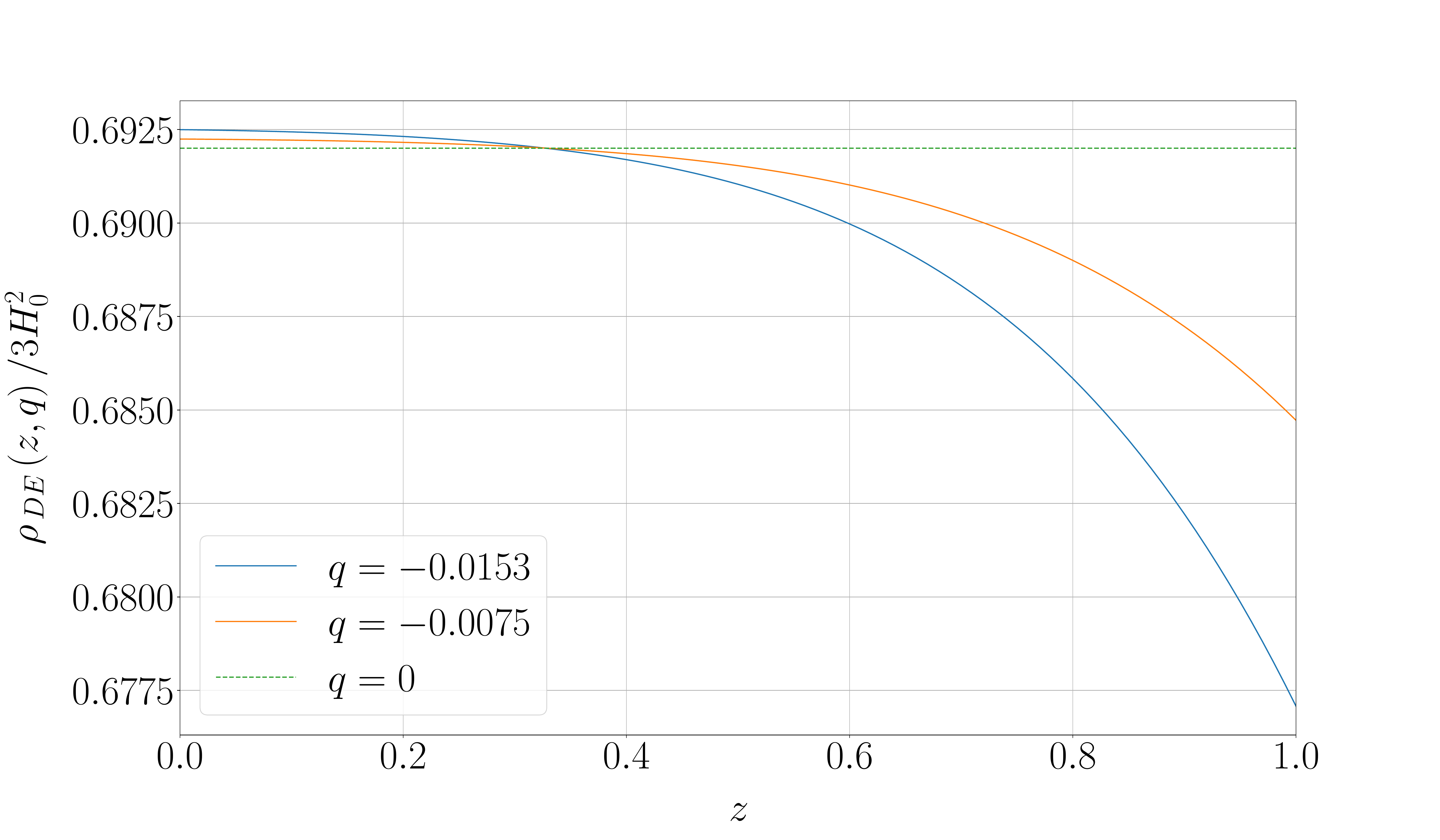}
\caption{The effective energy density from modified gravity, computed with Eq. (\ref{rhoDE}) (a), and its discrepancy $\Sigma$, as defined in Eq. \eqref{Sigma}, against the XCDM parametrization density Eq. \eqref{rhoXCDM} (b), in the case of $f(R,G)$-gravity.
\label{Fig8}}
\end{figure}

The behaviour of our model as a phantom fluid only can be seen in Fig.~\ref{Fig8}(a), where the effective $\rho_{DE}/(3H_0^2)$ is plotted. In Fig.~\ref{Fig8}(b) we plot the discrepancy on $\rho_{DE}$ between XCDM parametrization and the reconstrcuted modified gravity model; the error is smaller than $0.5\%$.

\subsubsection{Chevallier-Polarski-Linder parametrization.}

\begin{figure}[!h]
\includegraphics[width=0.5\textwidth]{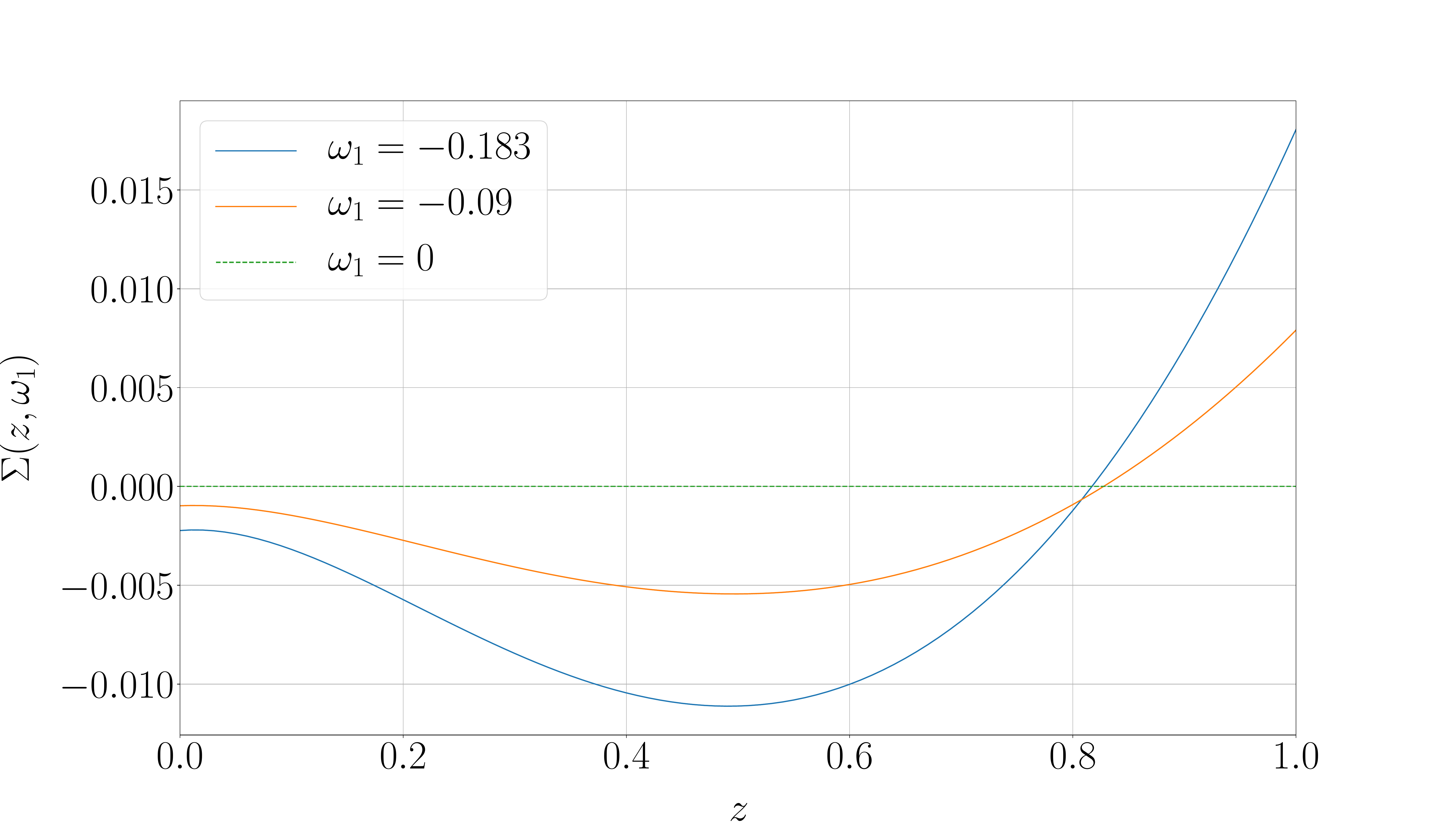}
\includegraphics[width=0.5\textwidth]{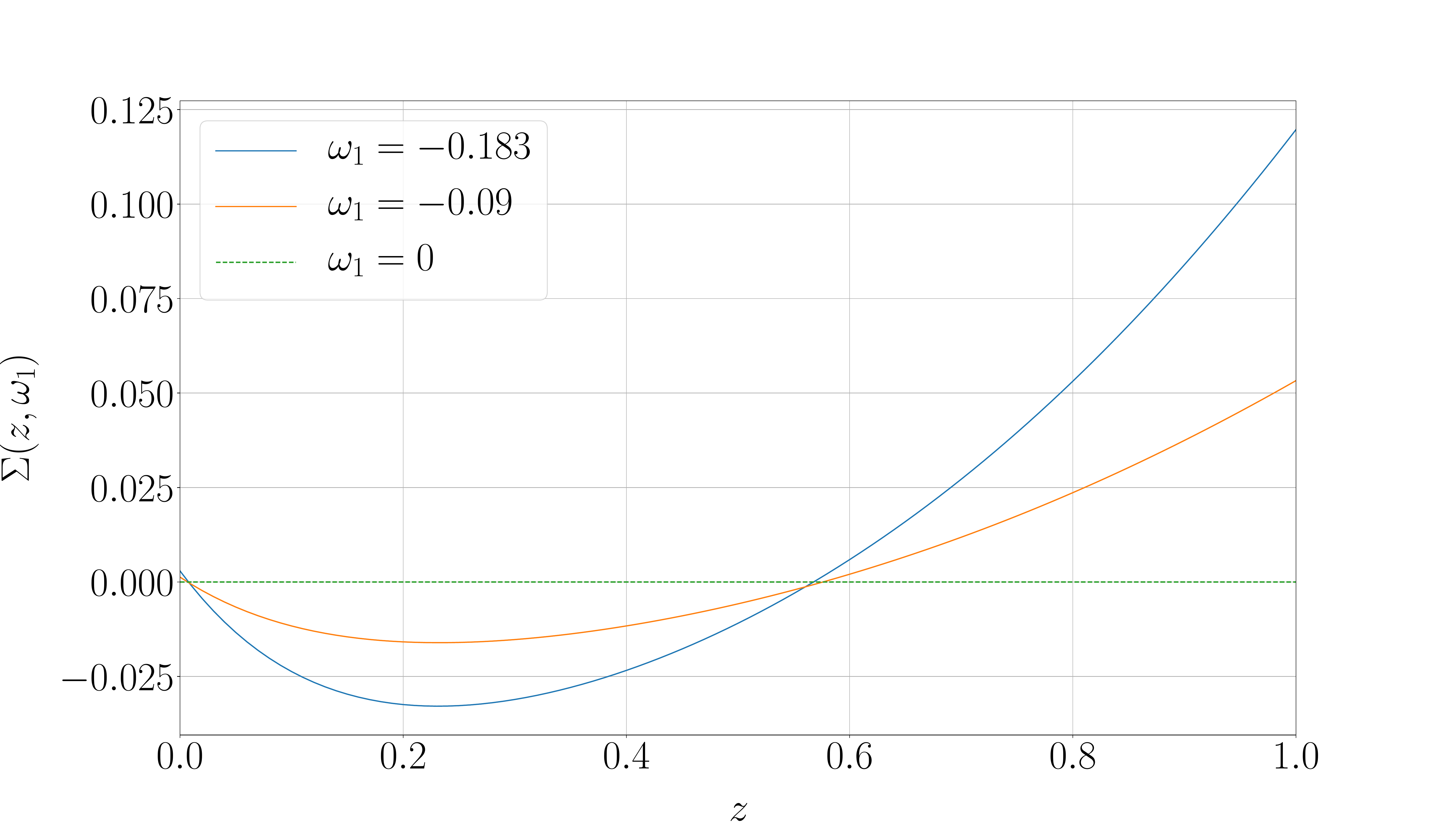}
\caption{Plots of the discrepancy function $\Sigma$, as defined in Eq. \eqref{Sigma}, with $J=H^2$ (a), and with $J=(z+1)H'/H$ (b), for the WP in the case of $f(R,G)$-gravity.
\label{Fig9}}
\end{figure}

In this section we consider the CPL parameterization, whose Hubble function is defined in Eq. (\ref{An2}). As in the case of $f(R)$, we fix $\omega_0=-1$ and we consider $\omega_1$ to be in the range as in Eq. (\ref{omegaCPL}). The corresponding $f(R,G)$-modified gravity function is derived in a numerically, using the proposal function
\begin{equation}
f(R, G)=-2\Lambda\left(1-{\frac{c_1\, \omega_1 X +c_2 \omega_1 (X^2/\Lambda)}{2\Lambda}}\right)\,,\quad X=\frac{G}{R}\,.
\end{equation}
The values of the parameters are
\begin{equation}
c_1=0.039\quad \text{and} \quad c_2=-0.011.
\end{equation}
As in the case of $f(R,G)$ XCMD parametetrization, the viable range of $\omega_1$ is restricted to the negative values only due to the physical constraints on $g(X)$ at step \eqref{step4} of the numerical reconstruction procedure. Therefore the parametrization parameter can have values in the range
\begin{equation}
-0.183<\omega_1<0\,.
\end{equation}
In Fig.s \ref{Fig9}(a) and \ref{Fig9}(b) we plot the discrepancies on $H^2$
and on $(z+1)H'/H$ between the CPL parametrization and the reconstructed modified gravity model. The discrepancy on the Hubble function is smaller than $2\%$, while on $(z+1)H'/H$ is smaller than $10\%$.

\begin{figure}[!h]
\includegraphics[width=0.5\textwidth]{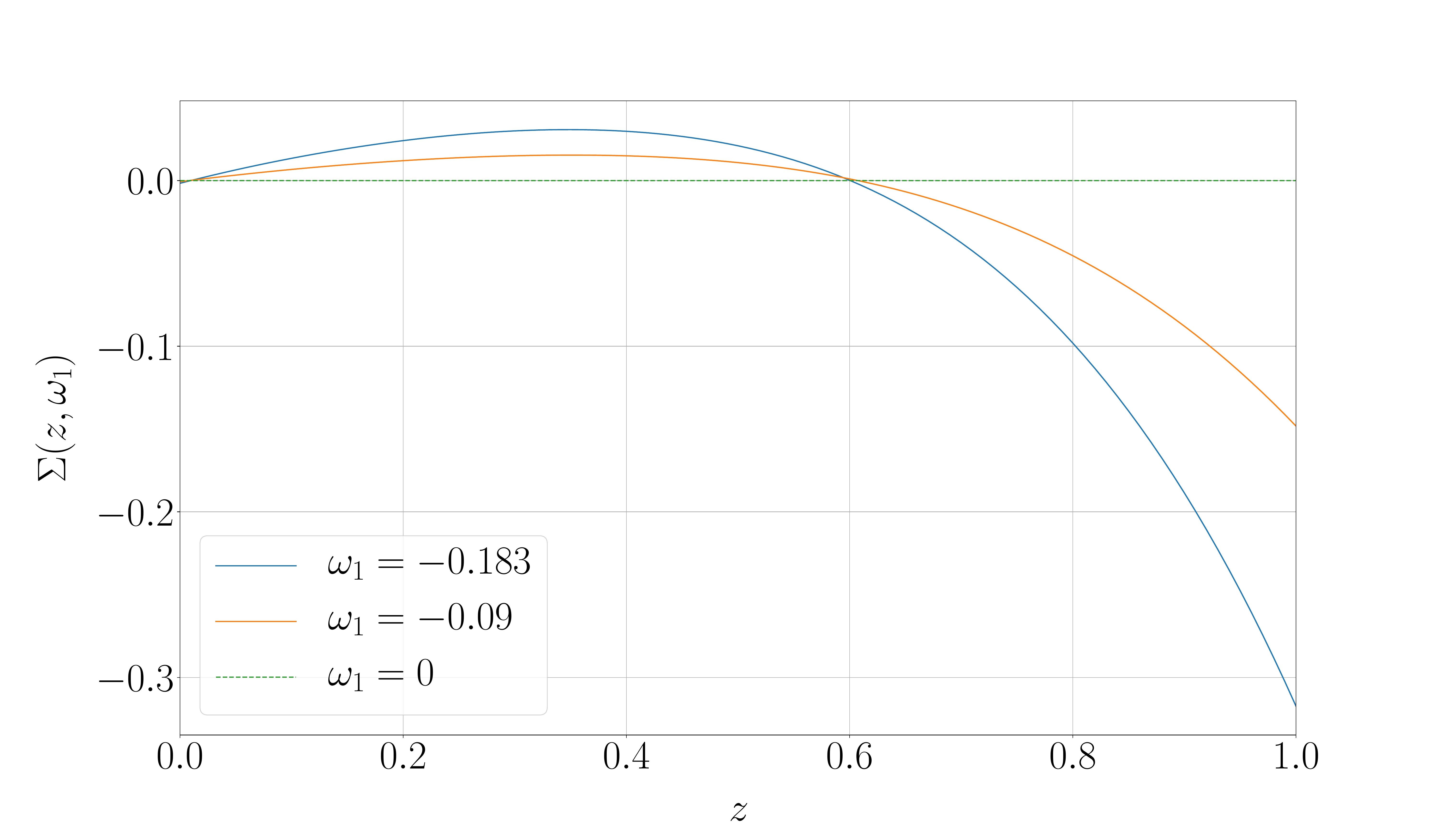}
\includegraphics[width=0.5\textwidth]{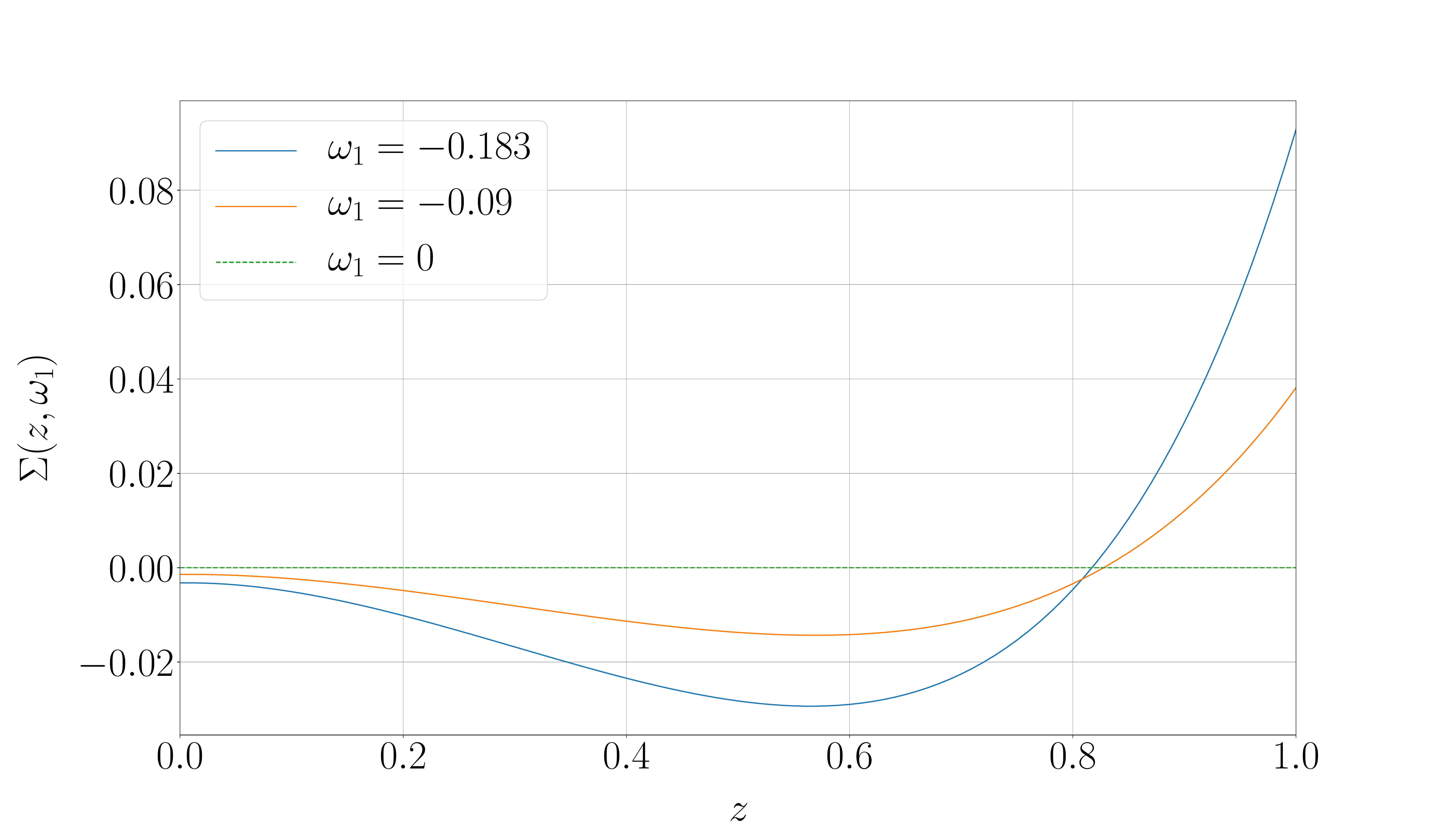}
\caption{The discrepancy $\Sigma$, as defined in Eq. \eqref{Sigma}, with $J=\rho_{DE}/(3H_0^2)$  (a) and with the equation of state parameter $J=\omega_{DE}$ (b), for the CPL parametrization in the case of $f(R,G)$-gravity.
\label{Fig10}}
\end{figure}

Finally, the discrepancies on $\rho_{DE}$ and on $\omega_{DE}$, between the CPL parametrization and the reconstructed modified gravity model, are shown in Fig.s \ref{Fig10}(a) and \ref{Fig10}(b) respectively.
The errors on the dark fluid energy density are smaller than $8\%$, while on the equation of state parameter reach the $25\%$ at high redshift for large (negative) values of $\omega_1$.

\subsubsection{Wetterich-redshift parametrization.}

\begin{figure}[!h]
\includegraphics[width=0.5\textwidth]{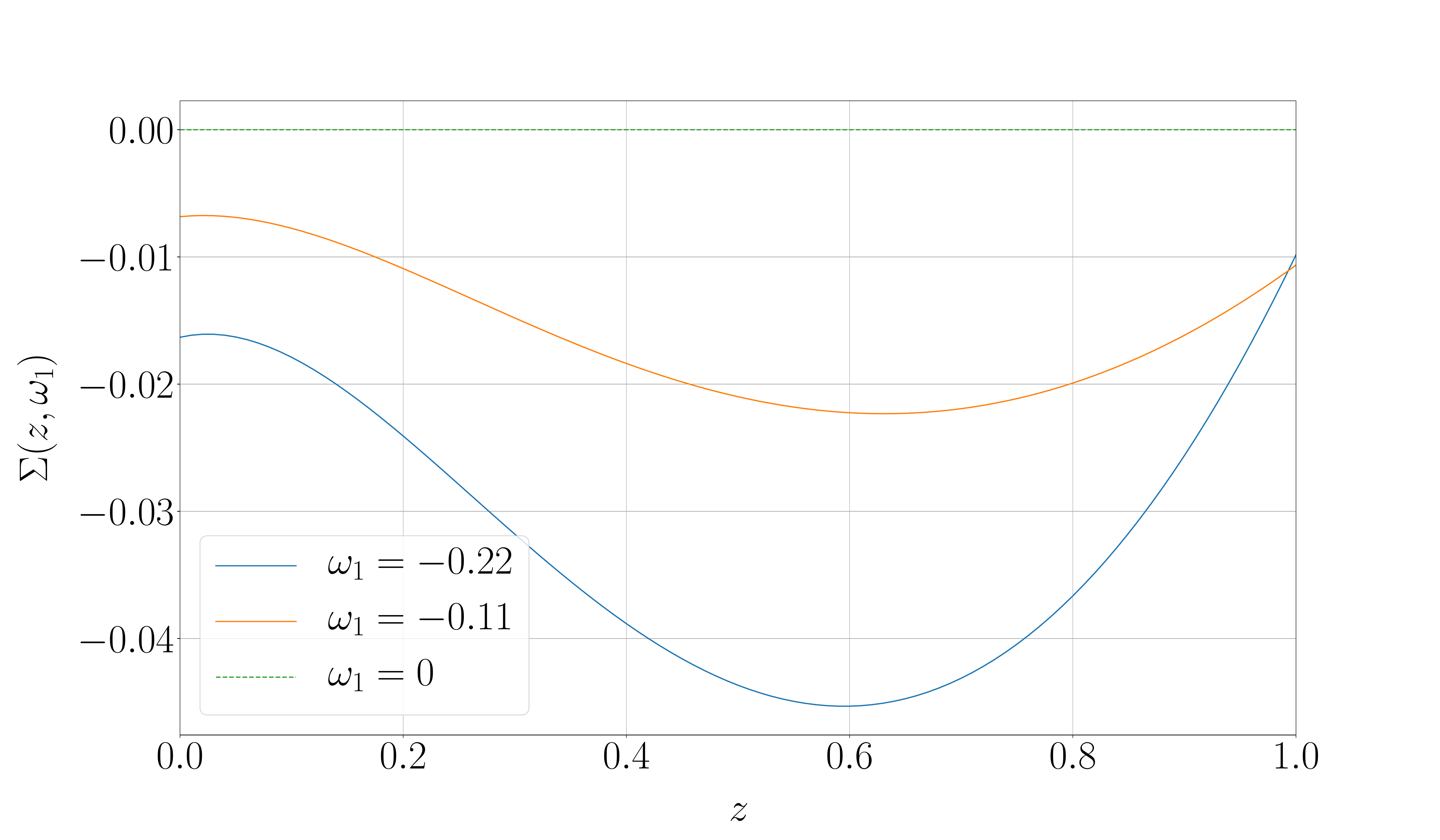}
\includegraphics[width=0.5\textwidth]{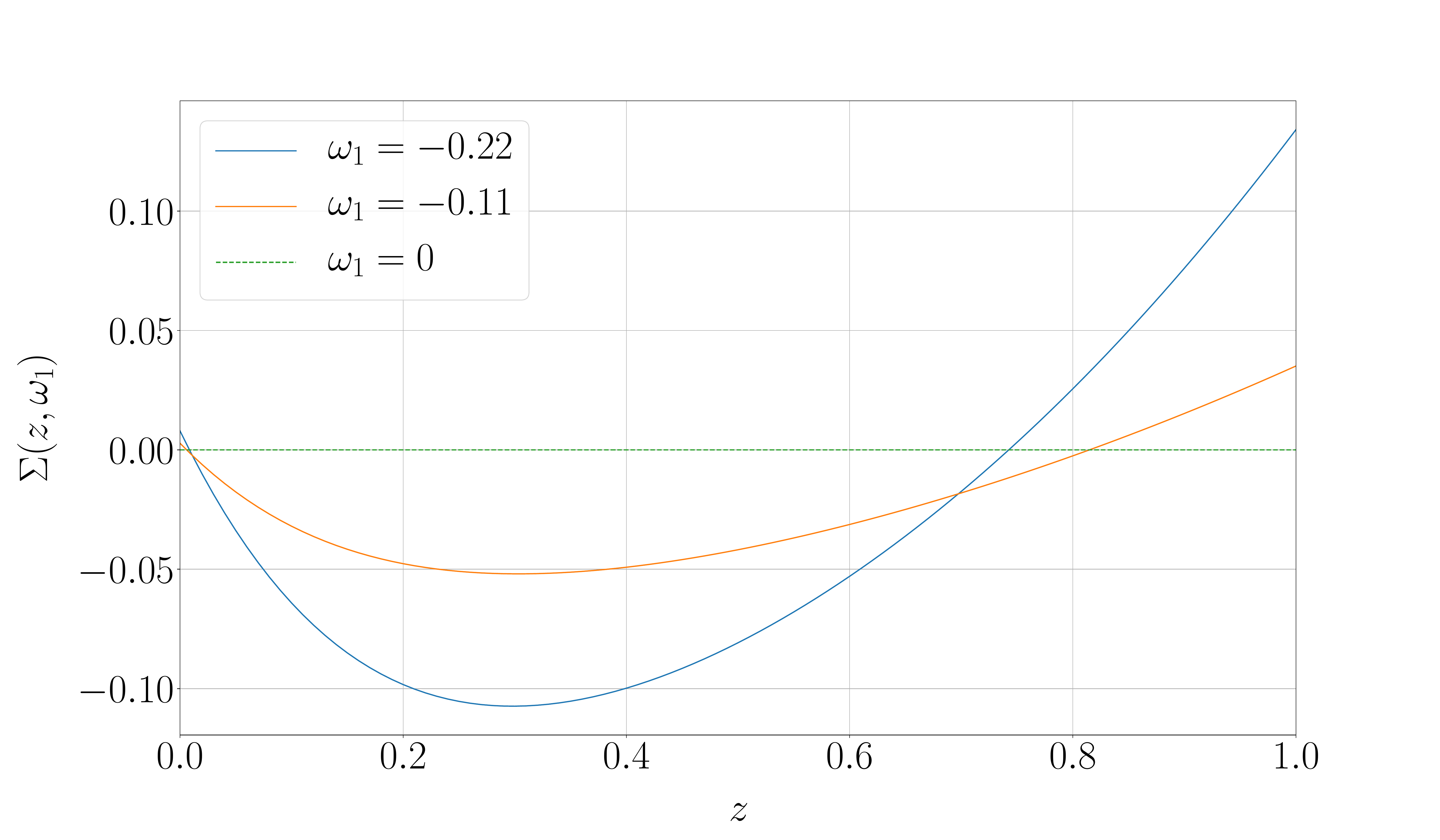}
\caption{The errors on $H^2$ (a) and on $(z+1)H'/H$ (b) in the WP.
\label{Fig11}}
\end{figure}

Finally, the last parametrization we present is the WP. The Hubble function for this parametrization is given by (\ref{An3}). Again we fix $\omega_0=-1$, while $\omega_1$ can take values in the range in Eq. (\ref{om1WP}). The proposal function we use is
\begin{equation}
f(R,G)=-2\Lambda\left(1-{\frac{c_1\, \omega_1 X +c_2 \omega_1 (X^2/\Lambda)}{2\Lambda}}\right)\,,\quad X=\frac{G}{R}\,.
\end{equation}
The values of the parameters are
\begin{equation}
c_1=0.15\,,\quad c_2=-0.0098.
\end{equation}
In order to avoid matter instabilities, similarly to the previous cases analysed in $f(R,G)$, we should impose the following condition on the parameter of the parametrization
\begin{equation}
-0.22<\omega_1<0.
\end{equation}
In Fig.s \ref{Fig11}(a) and \ref{Fig11}(b) we show the plots of the discrepancies on $H^2$
and on $(z+1)H'/H$ between the WP and the reconstructed modified gravity model. The error on the Hubble function is smaller than $5\%$, while on $(z+1)H'/H$ is smaller than $10\%$.

\begin{figure}[!h]
\includegraphics[width=0.5\textwidth]{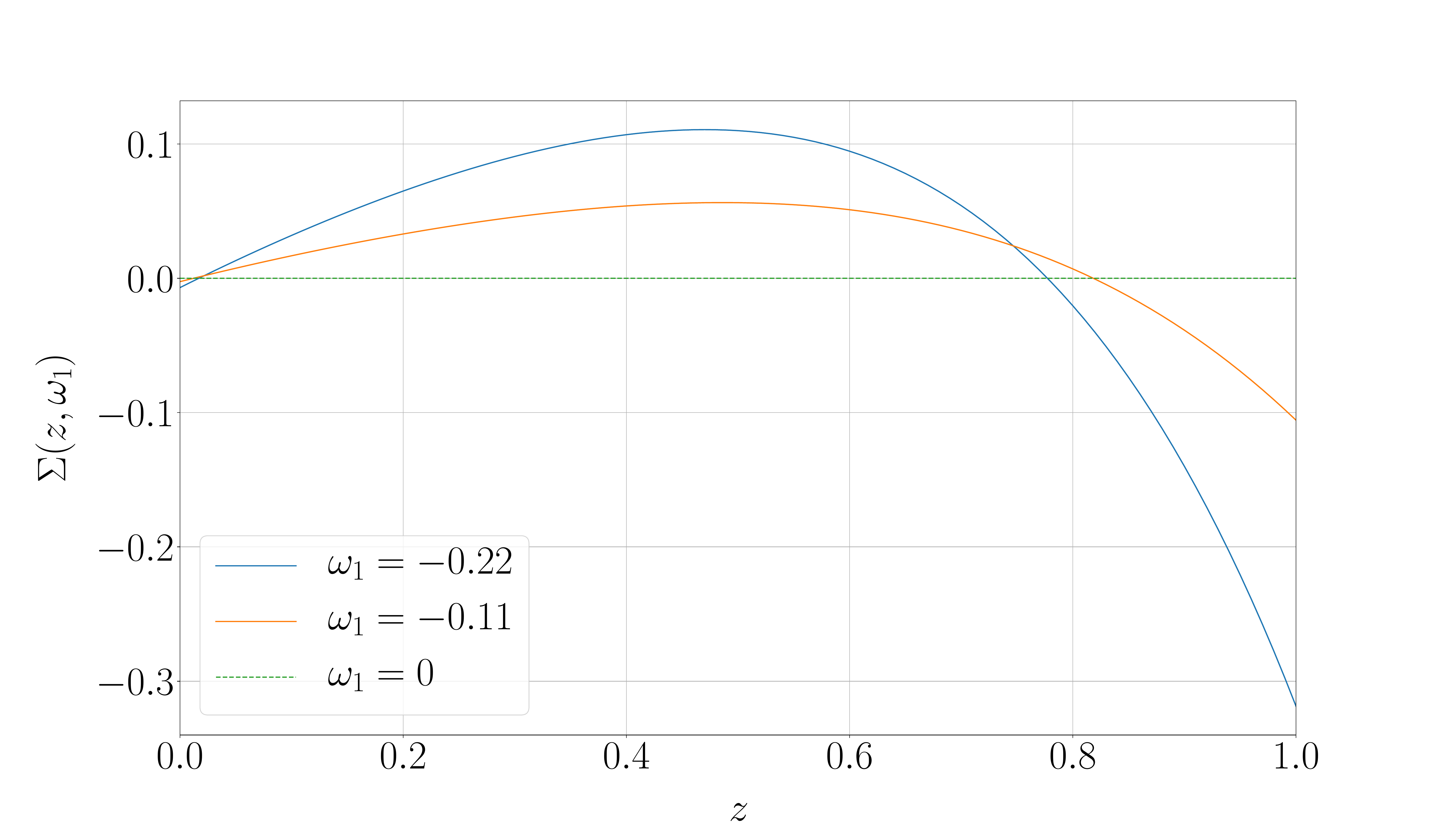}
\includegraphics[width=0.5\textwidth]{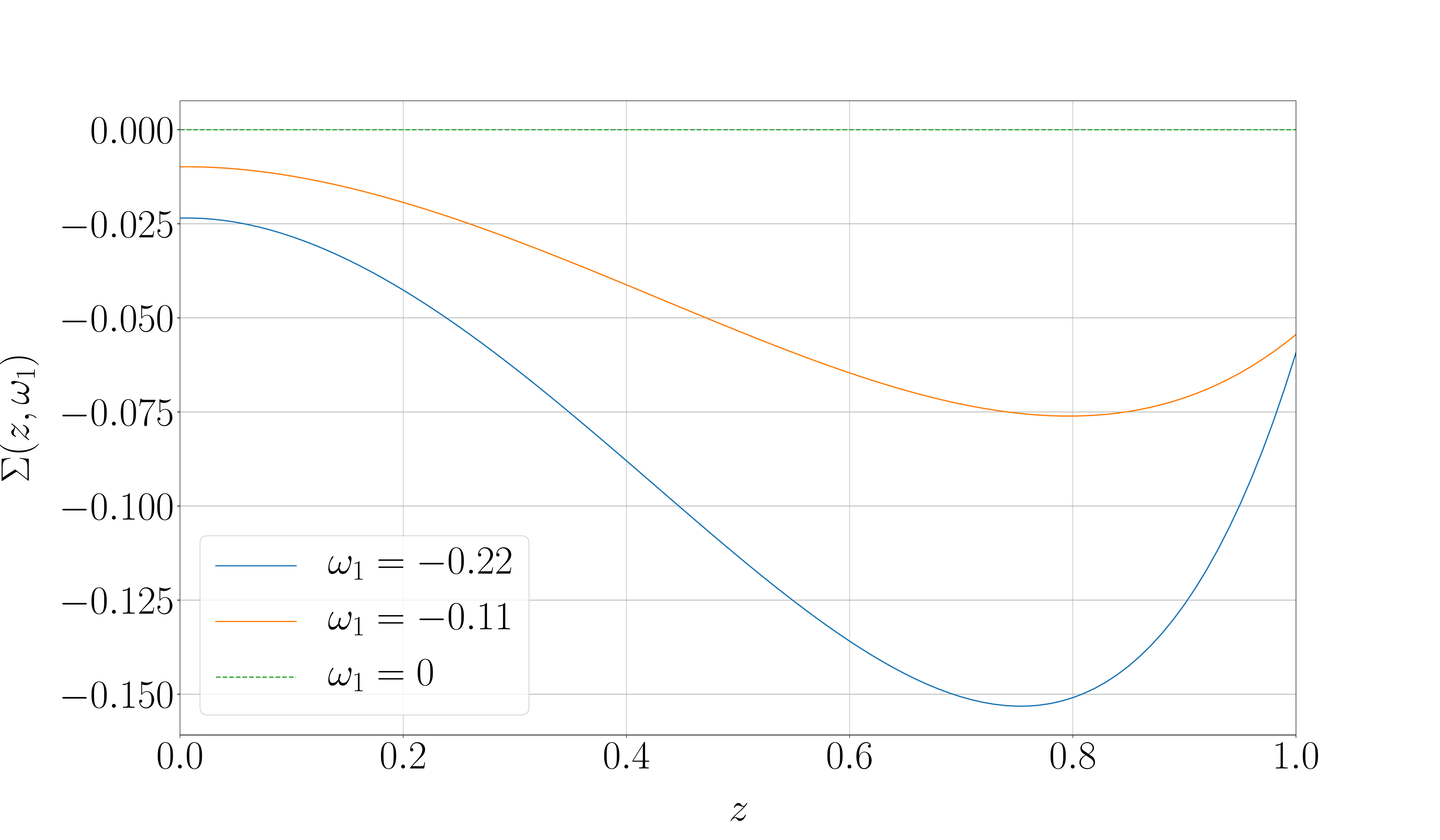}
\caption{The comparison of dark fluid from modified gravity and the perfect fluid description of WP (a). The comparison of $\omega_{DE}$ from modified gravity and the perfect fluid description of the WP (b).
\label{Fig12}}
\end{figure}

The discrepancy functions for $\rho_{DE}$ and $\omega_{DE}$ are shown in  Fig.s \ref{Fig12}(a) and \ref{Fig12}(b). While the error on $\rho_{DE}$ reaches the $15\%$, the error on $\omega_{DE}$ reaches the $30\%$ at high redshift and for large and negative values of $\omega_1$.

\section{Extrapolation at intermediate red-shifts for $f(R,G)$ and $f(R)$ theories}\label{sec:extrap}

The Taylor expansions we considered in the reconstruction procedure approximate $f(R)$ and $f(R,G)$ within the red-shift range $z \in [0,1]$ with sufficiently small errors. This procedure is jeopardized by the issue of truncated Taylor series, which turns out to be the exact reconstruction only if an infinite number of terms is taken into account. Since it is impossible to include such an infinite number of terms, one can wonder whether our extrapolated results can lead to a consistent expressions for higher red-shift data.
A simple approach to answer this question is to extrapolate functions in the form of $f(R,G)$ and $f(R)$ in a more general way, under the following requirements:

\begin{itemize}
  \item the new extrapolation should reduce to the approximations previously obtained by our methods;
  \item the numerical limits of the coefficients must be compatible with our previous reconstruction technique outcomes;
  \item the introduction of new parameters should not considerably complicate the whole statistics;
  \item since the extrapolations cannot be model-dependent, we should consider a model-independent expansion series.
\end{itemize}

The latter requirement is important since one does not have to fix \emph{a priori} the forms of additional coefficients, which remain unfixed at intermediate red-shift domains. However, even in this case, if one desires to avoid infinite numbers of terms, the series might be truncated at some finite order. This procedure clearly suffers from a severe divergence behaviour due to the truncation of the series, which influence the intermediate red-shift numerical limits .

Hence, in order to extend our work at intermediate red-shifts curvature and Gauss-Bonnet values, i.e. $z \in [0,2]$, without including higher orders of the series which may imply a broadening of the parameter value probability distribution, we can change our approach in favor of a new series definition which extends the previous one and fulfills the aforementioned requirements.

\begin{figure}[!h]
\centering
\includegraphics[width=0.4\textwidth]{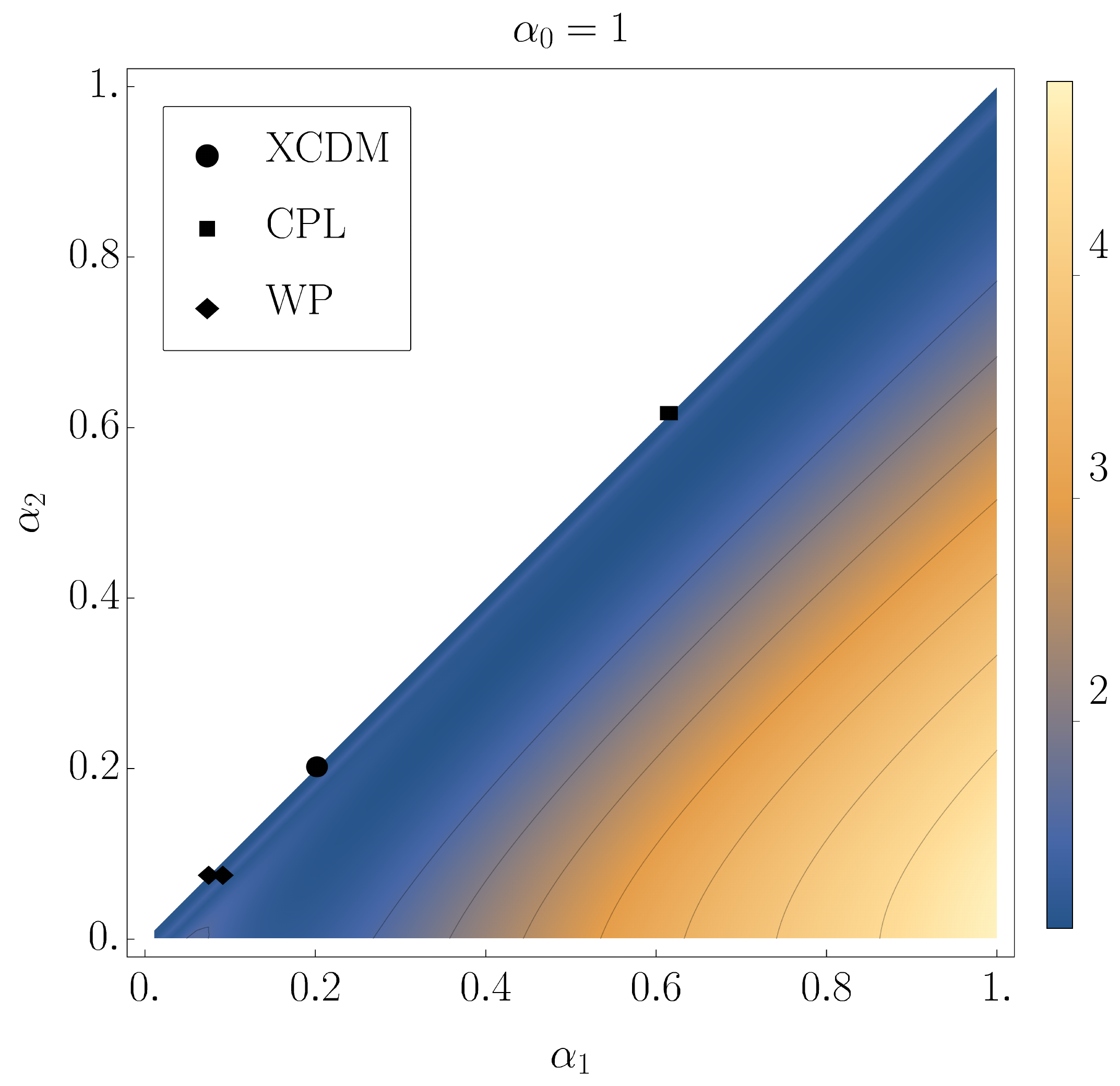}
\caption{Plot of the reduced $\chi^2$, $\tilde{\chi}^2$ as in equation \eqref{chi2}, computed with the data in Table \ref{tab:OHD} and with the solution $H(z)$ of the differential equation \eqref{equazioneRG0}, using the $f(R,G)=F(X)$ function \eqref{frg}, as the theoretical model. The white region is affected by heavy instabilities, and therefore has a $\tilde{\chi}^2\gg 1$. The points are the results shown in Table \ref{tab:results1}. Some points overlap, and therefore are not visible using this plot scale. Note that, in principle we should make a different plot for the points in Table \ref{tab:results1} which have a $\alpha_0\neq 1$. However the background $\tilde{\chi}^2$ does not change readily for small modifications of $\alpha_0$, and we can plot all points here for simplicity, in order to appreciate with one plot that all the point are in the blue, low $\tilde{\chi}^2$, region.
\label{fig:frg}}
\end{figure}

\subsection{The Pad\'e series}

In order to perform the extrapolation, a possibility which satisfies the above requirements is offered by the Pad\'e series for which it is valid
\begin{equation} \label{facy1}
  f(x)=P_{mn}(x)+ \sum_{i=1}^{\infty} c_{m+n+i} x^{m+n+i}  \,,
\end{equation}
where we define the $(m,n)$ order Pad\'e series as
\begin{equation} \label{facy2}
  P_{mn}(x) = \frac{ \sum_{i=0}^{m}a_i x^i}{1+\sum_{j=1}^{n}b_j x^j} \,,
\end{equation}
and $a_i,b_i$ are the free coefficients. The equivalence between the Taylor and Pad\'e series is guaranteed as one requires that at $x=0$ the two series reduce to the same outcome. Moreover, in order to get a balanced correspondence between the two series, the number of coefficients should be the same. The  main advantage of using the Pad\'e series is to obtain convergence at higher values of the variable $x$. In particular, the Pad\'e series has the following properties:

\begin{itemize}
  \item the $(m,n)$ orders should be equivalent to the order made by Taylor series in our previous approaches;
  \item the Pad\'e series is stable for a wider range of the variable $x$ with respect to the Taylor series;
  \item the Pad\'e series extrapolates the corresponding behavior of the Taylor series in a model-independent way;
  \item the Pad\'e series coincides with the Taylor series at $x=0$.
\end{itemize}

As a consequence, the Pad\'e formalism, besides small drawbacks, provides the great advantage to match intermediate domains, where conventional Taylor treatments fail to be predictive. Following the outlined treatment, we re-express the functions previously presented by fulfilling the basic demands of Pad\'e series. 

\subsection{Extrapolation procedure}

We propose the following strategy in order to extrapolate the values of the Pad\'e parameters:
\begin{itemize}
    \item we take the $(1,1)$ form of Pad\'e series in order to match the order of the Taylor series we got in the previous sections. Thus we use the extrapolation functions
    \begin{equation}\label{frg}
f(R,G)=-2\Lambda\left(\frac{\alpha_{0}+\alpha_{1}X/\Lambda}{1+\alpha_{2}X/\Lambda}\right)\,,
\end{equation}
and
\begin{equation}\label{fr}
f(R)=-2\Lambda\left(\frac{\beta_{0}+\beta_{1}R/\Lambda}{1+\beta_{2}R/\Lambda}\right)\,,
\end{equation}
where $\alpha_{0},\alpha_{1},\alpha_{2}$ for $f(R,G)$ and $\beta_{0},\beta_{1},\beta_{2}$ for $f(R)$ are constants;
\item we get the Hubble function evolution (up to $z=2$) solving the Friedmann equations \eqref{equazioneRG}\footnote{The differential equation \eqref{equazioneRG} is a third order differential equation for $H$ in the $z$ variable. Therefore, the Cauchy problem requires three initial conditions for $H(z=0) = H_0$, $\dot{H}(z=0) = \dot{H}_0$ and $\ddot{H}(z=0) = \ddot{H}_0$. The value of these initial conditions is in equations \eqref{thingsnow1}.} considering the extrapolation functions defined above;
    \item we fit the Hubble parameter function by means of differential ages from the catalog of Hubble data coming from the standard kinematics analyses of the supernova data in the range $z\in [0.781,1.965]$, shown in Tab. \ref{tab:OHD}. 
\end{itemize}

\begin{table}[h]
\setlength{\tabcolsep}{2em}
\begin{center}
\begin{tabular}{c c c }
\hline
\hline
 $z$ &$H \pm \sigma_H$ &  Ref. \\
\hline
0.781 & $105.0 \pm 12.0$ & \cite{Moresco12} \\
0.875 & $125.0 \pm 17.0 $ & \cite{Moresco12} \\
0.88	& $90.0 \pm 40.0$ & \cite{Stern10} \\
0.9 & $117.0 \pm 23.0$ & \cite{Simon05} \\
1.037 & $154.0 \pm 20.0$ & \cite{Moresco12} \\
1.3 & $168.0 \pm 17.0$ & \cite{Simon05} \\
1.363 & $160.0 \pm 33.6$ & \cite{Moresco15} \\
1.43	& $177.0 \pm18.0$ & \cite{Simon05} \\
1.53	& $140.0	\pm 14.0$ & \cite{Simon05} \\
1.75	 & $202.0 \pm 40.0$ & \cite{Simon05} \\
1.965& $186.5 \pm 50.4$ & \cite{Moresco15} \\
\hline
\hline
\end{tabular}
\caption{Differential age $H(z)$ data used in this work. The Hubble rate is given in units of km/s/Mpc.}
 \label{tab:OHD}
\end{center}
\end{table}

In the latter step of our procedure, we evaluate the goodness of the interpolation computing the reduced $\chi^2$, namely
\begin{equation}\label{chi2}
    \tilde{\chi}^2 = \frac{\chi^2}{N_\text{data}}= \frac{1}{N_\text{data}} \sum_{\text{ data}} \frac{\left[H_\text{data}(z)-H_\text{theo}(z)\right]^2}{\sigma_H^2}\,,
\end{equation}
where $N_\text{data}$ is the number of experimental degrees of freedom, that is the difference between the number of data available (the number of data shown in Table \ref{tab:OHD}) and the number of theoretical parameters (e.g. the $\alpha_i$); $H_\text{data}$ and $H_\text{theo}$ are respectively the Hubble function at a certain redshift coming from the data and the theoretical results computed as previously explained; $\sigma_H$ are the errors on the experimental data; the summation is over the all experimental data available. Our results will be more in accordance with the experimental data when the $\tilde{\chi}^2 \cong 1$.

We expect the results, which are independent to the reconstruction methods, to be compatible with the ones obtained from previous reconstruction analysis. In order to check the compatibility, we need to find the relation between the parameters $c_i$ of the reconstruction methods with the new independently computed $\alpha_i$ (and similarly for $f(R)$). We exploit the requirements of the Pad\'e series previously listed. In particular we use the requirement of having the same Taylor and Pad\'e series at $z=0$. Thus we require the derivatives of the extrapolation functions in the equations \eqref{frg} and \eqref{fr} to be equal to the derivatives of the functions used in the reconstruction procedure \eqref{Mod1_fRG} and \eqref{Mod1} respectively, at $z=0$. This procedure ensures that, at sufficiently low redshifts, the Taylor and the Pad\'e series are compatible. Since the order of the Pad\'e series considered is $(1,1)$, we apply this procedure up to the second order derivatives in the arguments $R$ or $X$ \cite{Capozziello:2019cav}. We obtain the following $q$-dependent relations for $f(R,G)=f(X)$
\begin{subequations}\label{AAA}
\begin{align}
    \alpha_0 &= \frac{2 c_1 \Lambda ^3+c_2^2 q X_0^3+6 c_2 \Lambda ^2 X_0}{2 c_1 \Lambda ^3+6 c_2 \Lambda ^2 X_0}\,,\\
    \alpha_1 &= -\frac{c_1^2 \Lambda ^2 q+3 c_1 c_2 \Lambda  q X_0+3 c_2^2 q X_0^2+2 c_2 \Lambda ^2}{2 c_1 \Lambda ^2+6 c_2 \Lambda  X_0}\,,\\
    \alpha_2 &= -\frac{c_2 \Lambda }{c_1 \Lambda +3 c_2 X_0}\,,
\end{align}
\end{subequations}
where $X_0 = X(z=0)$, and $q$ should be exchanged with the parameter of the Hubble parametrization considered in the reconstruction procedure (for instance, in the case of CPL parametrization, $q$ is $\omega_1$). In the case of $f(R)$ we obtain similar results: we simply need to substitute $c_2 \rightarrow q c_2$ in the $f(R,G)$ results since the form of the $f(X)$ functions \eqref{Mod1_fRG} and \eqref{Mod1} only differ for a $q$ factor. The consistency of the two procedures (the reconstruction procedure and the one with the Pad\'e series) is ensured by the same choice of the initial conditions for the value of $H(z=0)$ and $H'(z=0)$: therefore in both procedures, $X_0$ has the same value, computed in equation \eqref{thingsnow2}. If the points lies within the region of $\tilde{\chi}^2 \cong 1$, we can conclude that the two methods are compatible, since the Pad\'e series correctly extends (with respect to the available data) the reconstruction method results at higher redshifts, leaving the low-redshift behaviour untouched.

\subsection{Results}

We firstly present the values of the parameters computed with \eqref{AAA} using the values of $c_i$ and $q$ from the previous reconstruction method. They are available in Table \ref{tab:results1}. We note that the results are consistent with the $\Lambda$CDM limit of our procedure, since when the $q$ parameter is null, the $\alpha_i$ are such that $f(X)=-2\Lambda$ in all $q=0$ cases. Small modifications of $q$ from $0$ (within the limits discussed in the reconstruction method sections) give the non-trivial results. In the last column we show the value of the $\tilde{\chi}^2$ computed with the $\alpha_i$ values in the previous columns: in every case, the deviation from the $\Lambda$CDM case in small, confirming the compatibility of the two approaches (in the sense explained above).

\begin{table*}
\begin{center}
\setlength{\tabcolsep}{1em}
\renewcommand{\arraystretch}{1.8}
\begin{tabular}{|c| c c c| c c c|}
\hline
\hline
Model & Parametrization & $q$ or $\omega_1$ & $\beta_0$ or $\alpha_0$ & $\beta_1$ or $\alpha_1$ &  $\beta_2$ or $\alpha_2$ & $\tilde{\chi}^2$ \\
\hline\hline
$\Lambda$CDM & - & - & $1$ & $0$ & $0$ & 1.12\\
\hline\hline
$f(R)$& XCDM & $-0.0153$ & 1 &  $-0.008948$ & $-0.008257$ & 1.09\\
&  & $0$ & $1$ & $0$ & $0$&1.12\\
&  & $0.117$ & $0.9881$ & $0.4967$ & $0.4827$&1.08\\\hline
& CPL & $-0.183$ & $1.00004$ & $-0.01547$ & $-0.01160$ &1.10\\
&  & $0$ & $1$ & $0$ & $0$& 1.12\\
&  & $0.311$ & $0.9997$ & $0.03862$ & $0.03186$ & 1.07\\\hline
& WP & $-0.1$ & $1$ & $-0.007$ & $-3 \times 10^{-10}$ & 1.12\\
&  & $0$ & $1$ & $0$ & $0$& 1.12\\
& & $0.089$ & $1$ & $0.006675$ & $2 \times 10^{-10}$& 1.12\\\hline\hline
$f(R,G)$& XCDM & $-0.0153$ & $0.999994$ & $0.2024$ & $0.2016$ & 1.13\\
&  & $0$ & $1$ & $0.2016$ & $0.2016$ & 1.12\\\hline
& CPL & $-0.183$ & $0.9998$ & $0.6171$ & $0.6128$ & 1.09\\
&  & $0$ & $1$ & $0.6128$ & $0.6128$ & 1.12\\\hline
& WP & $-0.22$ & $0.99998$ & $0.09127$ & $0.07467$ & 1.28\\
& & $0$ & $1$ & $0.07467$ & $0.07467$ & 1.12\\
\hline
\hline
\end{tabular}
\caption{Table of the parameters of the functions \eqref{frg} and \eqref{fr}. The first result is obtained with a simple fit of the data with the usual Hubble function equation $H(z) = H_0 [\Omega_\Lambda + \Omega_m (1+z)^3]^{1/2}$. The other results are obtained using equations \eqref{AAA}, with the values of $c_i$ and $q$ from the previous reconstruction method. In the last column we show the value of the $\tilde{\chi}^2$, defined in equation \eqref{chi2}, computed with the $\alpha_i$ values in the previous columns.}
 \label{tab:results1}
\end{center}
\end{table*}

In Figure \ref{fig:frg} we present an alternative picture to present the results, in the case of $f(R,G)$. The background function is the reduced $\chi^2$ \eqref{chi2}. Values of $\alpha_1=\alpha_2$ (with $\alpha_0$) make the extrapolation function \eqref{frg} $f(R,G)=-2\Lambda$ ($\Lambda$CDM): around this straight line, $\tilde{\chi}^2\cong 1$. All points from Table \ref{tab:results1} are in the stable region, with $\tilde{\chi}^2 \cong 1$. This results ensure that using the parameters of $\alpha_i$ in Table \ref{tab:results1} we obtain a Pad\'e series that extends the reconstruction method functions at higher redshifts with a behaviour compatible with the experimental data considered.

\section{Final outlooks and perspectives}

We considered two extensions of General Relativity, $f(R)$ and $f(R,G)$ theories. We analyze possible breakdowns of the standard $\Lambda$CDM paradigm at intermediate and small redshift. To reconstruct the functional forms of $f(R)$ and $f(R,G)$ models we assumed three viable parameterized $H(z)$ which correspond to effective dark energy fluids. Afterwards, we inverted $R$ and $X$ in terms of the redshift $z$, having the corresponding functions $z=z(R)$ and $z=z(X)$, in which $X$ is fixed to a particular invertible choice between $G$ and $R$, namely $X=\frac{G}{R}$. We thus limited our treatment by only assuming the concordance paradigm is preserved at small redshifts and by involving a few  classes of $F(R,G)=R+f(R,G)$ and $F(R)=R+f(R)$ models. Thus, by calibrating the shapes of our curves through  XCDM, CPL, WP dark energy models we fixed the values and the forms of the free parameters. The corresponding $f(z)$ auxiliary functions in which $z=z(R)$ and $z=z(X)$ have been discussed together with the phase space in which they are thought to be available. The coarse-grained inverse scattering procedure has been computed even in terms of discrepancies over the shapes of the functions. To do so, we evaluated the error propagation over the two functions got in our analysis. The corresponding 3D plots with the whole phase space indicated which regions are favored with respect to others. Once numerically reconstructed, the shapes  of $f(z)$ and of $H(z)$, we discussed the consequences of our approach within observable and theoretical cosmology in a wider range of redshifts. To do so, we considered the Hubble measurements and we extrapolate the shapes of curves from small to intermediate redshifts. The procedure has been carried out by means of rational approximations which are stable at high $R$ and $X$. Hence, by making use of the Pad\'e series we inferred the new functions and we matched the correspondence between these new shapes with the previous ones, i.e the functions at small redshifts. To check the validity of our choices, we compared the results with the Hubble measurements at redshift $z>0.75$ and we performed an analysis with a combined data sets in which we made use of small and intermediate catalogs of data. We finally compared the so-obtained functions and we checked the best corrections to Einstein's gravity using statistical criteria. 

At low redshifts we evaluated the discrepancies among the Hubble function and its derivative and our reconstructed modified gravity models. These discrepancies might be small and, in this respect, our method seems to disagree in the case of $f(R)$ parameterized by means of the  WP parametrization as $\omega_1 \lesssim -0.1$, whereas in all the other cases, the discrepancies are: $<6\%$ for the Hubble function and $<10\%$ for its derivative. In our picture, we stress the fact that positive $q$ and $\omega_1$ terms, in the $f(R,G)$ theories, are excluded to avoid matter instabilities. We found the link between the intermediate and low redshift functions to be compatible with the Hubble function experimental data. Although our procedure agreed with the simplest extensions of general relativity in terms of $R$ and $G$, it enables one to consider as most suitable $f(R)$ and $f(R,G)$ functions the are rational Pad\'e polynomials of (1,1) orders. These approximations seemed to agree with both numerical reconstructions at small and higher redshifts.

In future developments, we will show whether our solutions are suitable at higher redshifts and which corrections are expected in such a case. Furthermore, we will employ small perturbations to see the consequences of our approach to structure formations. We will analyse also how to reconcile high and small redshift using our approach.

\section*{Acknowledgements}
OL acknowledges INFN, Frascati National Laboratories, for Iniziative Specifiche MOONLIGHT2 for support. This article is also supported in part by the Ministry of Education and Science of the Republic of Kazakhstan by the Program "Fundamental and applied studies in related fields of physics of terrestrial, near-earth and atmospheric processes and their practical application" IRN: BR05236494.

\end{document}